\documentclass[5p]{elsarticle}

\usepackage{lineno,hyperref}
\usepackage{svg}
\usepackage{subfloat}

\modulolinenumbers[5]

\journal{Journal of \LaTeX\ Templates}

\usepackage{supertabular}
\usepackage{rotating}
\usepackage{float}
\usepackage[utf8x]{inputenc}
\newsavebox{\largestimage}
\usepackage{siunitx}
\usepackage{multirow}

\usepackage{multibib}
\usepackage{subfig}

\usepackage{array}
\newcolumntype{C}[1]{>{\centering\arraybackslash}p{#1}}

\usepackage{graphicx}
\usepackage{mwe}

\usepackage{numcompress}\biboptions{authoryear}

\usepackage{filecontents}
\usepackage{amsmath}
\usepackage{textcomp}
\newcommand{\corrb}[1]{\color{black}#1\color{black}}
\newcommand{\corrbb}[1]{\color{black}#1\color{black}}

\begin{document}

\begin{frontmatter}

\title{Fireball characteristics derivable from acoustic data}

\author[1]{Luke McFadden}
\author[1,2]{Peter Brown \fnref{myfootnote}}
\author[1]{Denis Vida}
\author[3]{Pavel Spurný}
\fntext[myfootnote]{Correspondence to: Peter Brown (pbrown@uwo.ca)}
\address[1]{Dept. of Physics and Astronomy, University of Western Ontario, London, Ontario, Canada N6A 3K7}
\address[2]{Centre for Planetary Science and Exploration, University of Western Ontario, London, Ontario, Canada N6A 5B7}
\address[3]{Astronomical Institute of the Czech Academy of Sciences, Ondřejov, Czechia}

\begin{abstract}
Near field acoustical signals from fireballs (ranges $<$ 200 km), when detected by dense ground networks, may be used to estimate the orientation of the trajectory of a fireball \citep{Pujol2005} as well as fragmentation locations \citep{Kalenda2014, Edwards2004}. Distinguishing ballistic arrivals (from the cylindrical shock of the fireball) from fragmentation generated signals (quasi-spherical sources) remains a challenge, but are obtainable through analysis of the acoustic path and the timing observed at ground instruments. Here we describe an integrated computer code, termed the Bolide Acoustic Modelling program or \texttt{BAM}, to estimate fireball trajectories and energetics.  We develop a new methodology for measuring energy release from bolide fragmentation episodes solely from acoustic measurements and incorporate this into \texttt{BAM}.  We also  explore the sensitivity of seismo-acoustic fireball solutions and energy estimates to uncertainty in the underlying atmospheric model. Applying \texttt{BAM} to the Stubenberg meteorite producing fireball, we find the total fireball energy from ballistic arrivals to be approximately $5 \times 10^{10}$ J which compares favorably to the optical estimate of $4.36 \times 10^{10}$ J. The combined fragmentation energy of the Stubenberg event from acoustic data was found to be $1.47^{+0.28}_{-0.12} \times 10^{10}$ J, roughly one third of the ballistic or optical total energy. We also show that measuring fireball velocities from acoustic data alone is very challenging but may be possible for slow, deeply penetrating fireballs with shallow entry angles occurring over dense seismic/infrasound networks.

\end{abstract}


\end{frontmatter}

\section{Introduction}
\subsection{Infrasound and Meteors}
Infrasonics is the study of sound waves with a frequency below the range of human hearing, from $20$ Hz down to the order of $10^{-3}$ Hz, or infrasound \citep{PichonBook2, Edwards2008}. \corrb{Such low-frequency sounds can be produced by earthquakes, avalanches, meteors, and nuclear explosions}. Infrasound is able to propagate far distances from the source before dissipation, much farther than audible sounds making it a useful tool for detection of various geophysical phenomena. Because of its low attenuation, infrasound may be detectable over large (global-scale) ranges and often provides information about its source otherwise unobtainable.\par
Among the geophysical sources of infrasound are large meteors, also called bolides or fireballs \citep{Silber2019}. When a meteoroid enters the atmosphere, it does so travelling much faster than the local atmospheric speed of sound. The strong shock produced at lower heights by larger meteoroids when interacting with the atmosphere creates a Mach cone: a near cylindrical cavity which propagates outward at supersonic speeds (initially) from the meteoroid path. For meteoroids moving with speeds ranging from Mach 35 to Mach 240 \citep{Edwards2008}, the Mach cone will have a small angle, and can be effectively approximated as a line source \citep{Revelle1975}. In this picture, the deposition of energy by the meteoroid produces a blast with a cylindrical shock geometry moving radially outward, which then decays at long ranges to a nearly linear acoustic wave (infrasound). All meteoroid entries produce such cylindrical (or ballistic) shocks, but the intensity and prominence of the shock when detected by ground sensors depends on many factors such as energy deposition at the source height, range and source height \citep{Revelle1976}.  \par
In addition to the cylindrical (or ballistic) shock produced by all meteoroids traversing the atmosphere, a meteoroid may also fragment during its flight \citep{Ceplecha1998}. This fragmentation produces a sudden increase in the rate of energy deposition and results in a quasi-spherical shock, independent of the Mach cone. \par 
All components of the low-frequency sound produced by the fireball representing the shock wave decaying at long ranges from the trajectory can be measured as infrasound on seismographs (when the sound couples to the solid earth) or directly by infrasonic microphones, provided the amplitude of the waves is sufficient to be detected \citep{Edwards2008}.\par
The meteoroid passage also produces optical luminosity, termed a fireball. If multiple cameras detect the same fireball and the light intensity recorded, the fireball velocity, orbit and energy may be computed. This is the most traditional means of gathering physical information about fireballs \citep{Ceplecha1998}. The light intensity as a function of time is a proxy for the energy deposited along the trajectory which in turn can be used to estimate shock production.\par 

Through the shock produced by the meteoroid, infrasound provides another method of measuring parameters of meteoroid trajectories, either supplementing or replacing optical observations. Using a sufficiently dense seismic and/or infrasound network, with stations close to the fireball ground path, the trajectory, time of appearance, total energy and fragmentation energies of a fireball can, in principle, be found. In practice this is often a very difficult task. The accuracy of acoustic trajectory reconstruction and energy estimation depends on several factors (see \citealp{Silber2019} for a detailed review). Among these factors are:
\begin{enumerate}
    \item The accuracy of the atmosphere/wind field used in ray tracing
    \corrb{\item The propagation method used in ray-tracing}
    \item The ability to distinguish ballistic from fragmentation-produced acoustic arrivals at any given station
    \item The applicability of analytic energy estimation techniques for both ballistic and fragmentation shocks
    \item Fireball trajectory orientation and position with respect to the seismic network
    \item The ability to measure fireball speed from acoustic arrivals alone.
\end{enumerate} 
\corrb{The last of these items (fireball speed) is generally regarded as essentially unconstrained by acoustic arrivals \citep[e.g.][]{Pujol2005} simply because of the large contrast between the meteoroid speed and atmospheric sound speed}. Speeds may be computed from single station optical measurements if the fireball geometry is known from acoustic observations. However, in some limited geometrical circumstances, such as meteors with grazing entry, velocity estimates for fireballs from acoustic data alone may be possible, as discussed later in Section \ref{discussion}. \par

Here we review past techniques developed to estimate fireball trajectories and geolocate fragmentation points. We implement these approaches into a single computer program, called the Bolide Acoustic Modelling (\texttt{BAM}) package, \corrb{which we describe in \ref{BAMdocs}}. We explore the uncertainty in derived quantities (fireball trajectory, fragmentation location, energy) associated with the known variance in the underlying atmospheric models. We also present a new approach for estimating fragmentation energies for fireballs directly from near-field fireball acoustics. Validation of these techniques, both kinematic and energetic, is made through comparison with the well known parameters of a fireball, observed  using optical and radiometric techniques, associated with the Stubenberg meteorite \citep{Spurny2016} fall, which occurred over a region with dense seismic and infrasound station coverage, making it an ideal case study.

\section{Theoretical Background}
\subsection{Types of Acoustic Arrivals (Ballistic vs. Fragmentation)}

In earlier studies, it has been shown that the time of arrivals of a fireball-produced shock wave at numerous well positioned infrasound and/or seismic stations can be used to find its fragmentation points \citep{Edwards2003}, as well as uniquely estimate the fireball flight path \citep{Pujol2005, Ishihara2003}. However, this is only possible if the acoustic returns associated with these two distinct types of shocks can be clearly separated in seismic or infrasound time series. These approaches have been verified through comparison with other instruments, usually optical cameras \citep[e.g.][]{Ishihara2004}.  However, as near-field (ranges less than 150 km) acoustic arrivals from fireballs are frequently complex wavetrains, distinguishing common fragmentation points from ballistic arrivals is demanding \citep{Tatum2000}, in the absence of an independent estimate for the fireball trajectory. This is one of the major challenges in fireball acoustic analysis and a major focus of our efforts. \par

When a meteoroid fragments, the resulting shock can be treated as a quasi-spherical shock wave, as discussed by \cite{Edwards2003}. This spherical source is normally considered to be a point source (rather than an extended source) and the problem becomes similar to that of earthquake geolocation, in this case the source being a point in the atmosphere \citep{Walker2011}. At ranges beyond 150-200 km, multiple ray paths cause distinct arrivals, which are difficult to distinguish without detailed ray tracing. Moreover, as range increases, the error in location increases due to uncertainty in the effective atmospheric sound speed; as a result, fragmentation geolocation is best accomplished using a selection of stations with short range to a particular fireball \citep{Hedlin2010}. \par
The techniques used to locate the ground position and height for fireball fragmentation points are well described in the literature \citep{Edwards2003, Anglin1988, Cumming1989, Qamar1995, Ishihara2004}, again presuming acoustic arrivals are uniquely identifiable with specific fragmentation points \citep{Kalenda2014}. We build on the algorithm, termed \texttt{SUPRACENTER}, presented by \cite{Edwards2003} for fragmentation localization. Figure \ref{frag_model} shows the model used to ray-trace from a fragmentation point, termed a ``supracenter''.\par

Similarly, the nonlinear inversion process to estimate a fireball trajectory given the timing of the ballistic shock arrivals at ground stations have been described in several earlier works \citep{Qamar1995, Tatum2000, pichon2002, Ishihara2004, Langston2004, Pujol2005}. Here we use the algorithm proposed by \cite{Pujol2005} modified to use the tau-p raytracing approach summarized in \cite{Garces1998}, which uses a change in variables (delay time $\tau$ and ray parameter $p$ instead of generic time and space variables \citep{Buland1983}) for a more computationally efficient algorithm. \corrb{We adapt the tau-p method taken from \citep{Edwards2003} into our Python package, to be consistent with their results and since it has been shown previously to work well in acoustic geolocation of fireball fragmentation points. } Figure \ref{ballistic_model} shows the parameters used to invert for a trajectory solution. \par 
\par

\begin{subfigures}
\begin{figure}[]{}
     \centering
     \includegraphics[width=\linewidth]{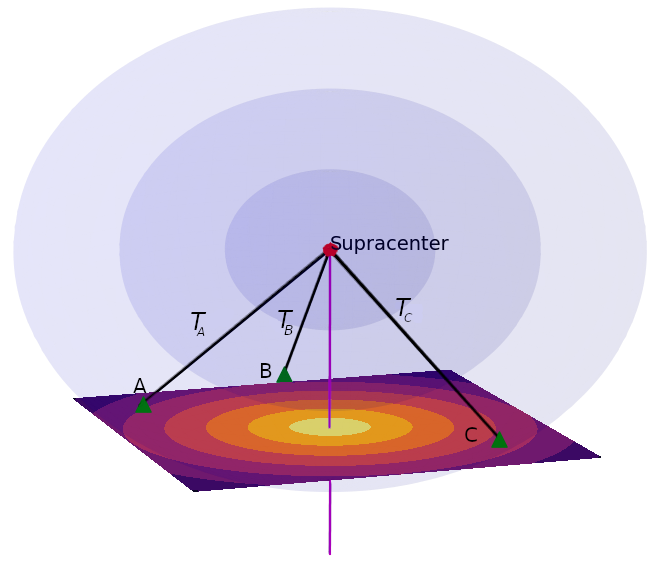}
    
    \caption{The ray-tracing of an atmospheric supracenter in 3-D space, shown for an isotropic atmosphere for simplicity. The point marked ``Supracenter" represents a fragmentation point on a fireball, and triangles $A$, $B$, and $C$ are seismic/infrasound stations. The contour on the ground shows the relative arrival times of the acoustic waves, with lighter representing earlier arrivals. The ray-tracing lines, $T$, represent the time it takes for the acoustic wave to reach each station. In reality, the atmosphere has winds and temperature gradients, and therefore, the $T$ lines are curved.}
    \label{frag_model}
\end{figure}
\begin{figure}[]

         \centering
    \includegraphics[width=\linewidth]{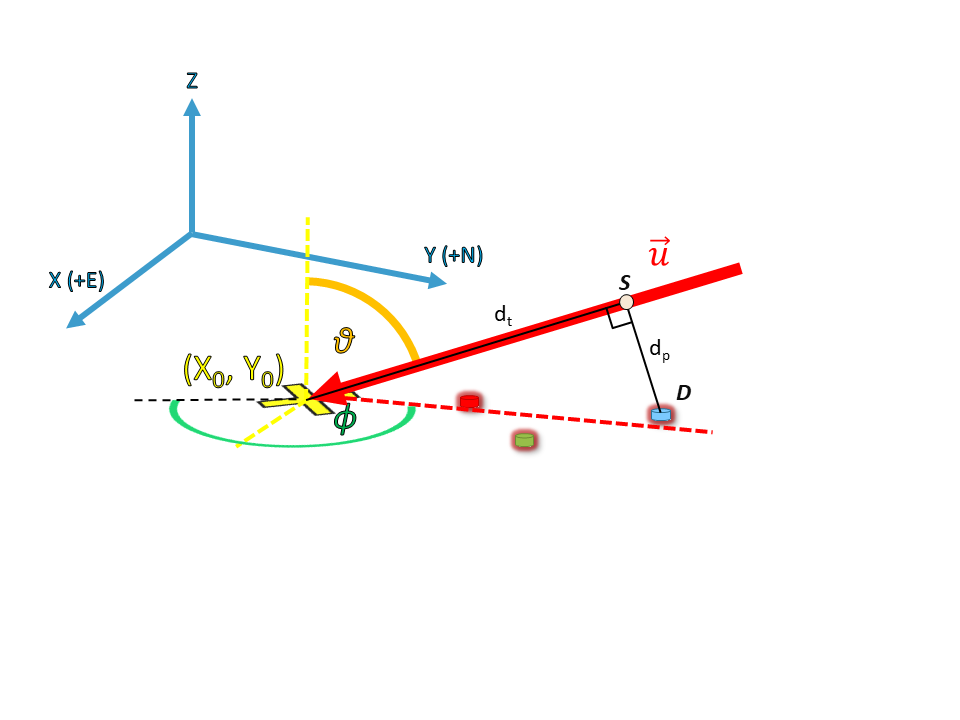}
    \caption{The ray-tracing of a meteor trajectory in 3-D space, shown for an isotropic atmosphere for simplicity. The local coordinate system, $(X, Y, Z)$ is described with $+Y$ pointing north and $+X$ pointing east. $(X_0, Y_0)$ represents the geometric landing point at $Z=0$, i.e.  where the extended fireball trajectory vector ($\vec{u}$ red line) intersects the ground. $\theta$ and $\phi$ are the zenith and azimuth angles of the trajectory vector, with the azimuth beginning at the north, and increasing towards the east as seen from above, defining the direction the meteor is heading towards. Points along the trajectory $\vec{u}$, may have acoustic paths to each station in a realistic atmosphere which we estimate through ray-tracing. The acoustic travel time is then calculated from time $t_0$, the time the trajectory would intersect the geometric landing point if it was travelling at a constant velocity, $v$. \corrbb{$D$ represents the coordinates of a specific station, and $S$ is the wave release point. The distance along the trajectory to the wave release point is $d_t$ while $d_p$ is the path of an acoustic ray from the wave release point to a specific station (further discussed in Figure \ref{ballisticdiagram})}. In reality, the ray-trace path, $d_p$ is curved, since the atmosphere has winds and temperature gradients.}
    \label{ballistic_model}
 \end{figure}
\end{subfigures}

    



While acoustic fireball trajectory solutions have been presented in the literature, the uncertainty and sensitivity of acoustically determined fireball solutions has been less well studied \citep[e.g.][]{Walker2011}. A fundamental limitation in localization of fireball trajectory or fragmentation points is the accuracy of the underlying atmosphere model used for ray-tracing and hence the effective sound speed (including winds) for propagation. This is another focal point we address in our work. 

\subsection{Effects of the Atmosphere}
The acoustic ground footprint of the fireball is heavily modified both by the atmospheric temperature structure and atmospheric winds. In an isothermal atmosphere, fireball ballistic (or cylindrical) shock produces a parabolic-shaped ground footprint while a fragmentation results in a spherical ground acoustic footprint. These simple, but distinct acoustic footprints, are shown in Figures \ref{ball_no_wind} and \ref{frag_no_wind} respectively. 

\begin{subfigures}
\begin{figure}[]
\centering

         \centering
         \includegraphics[width=\linewidth]{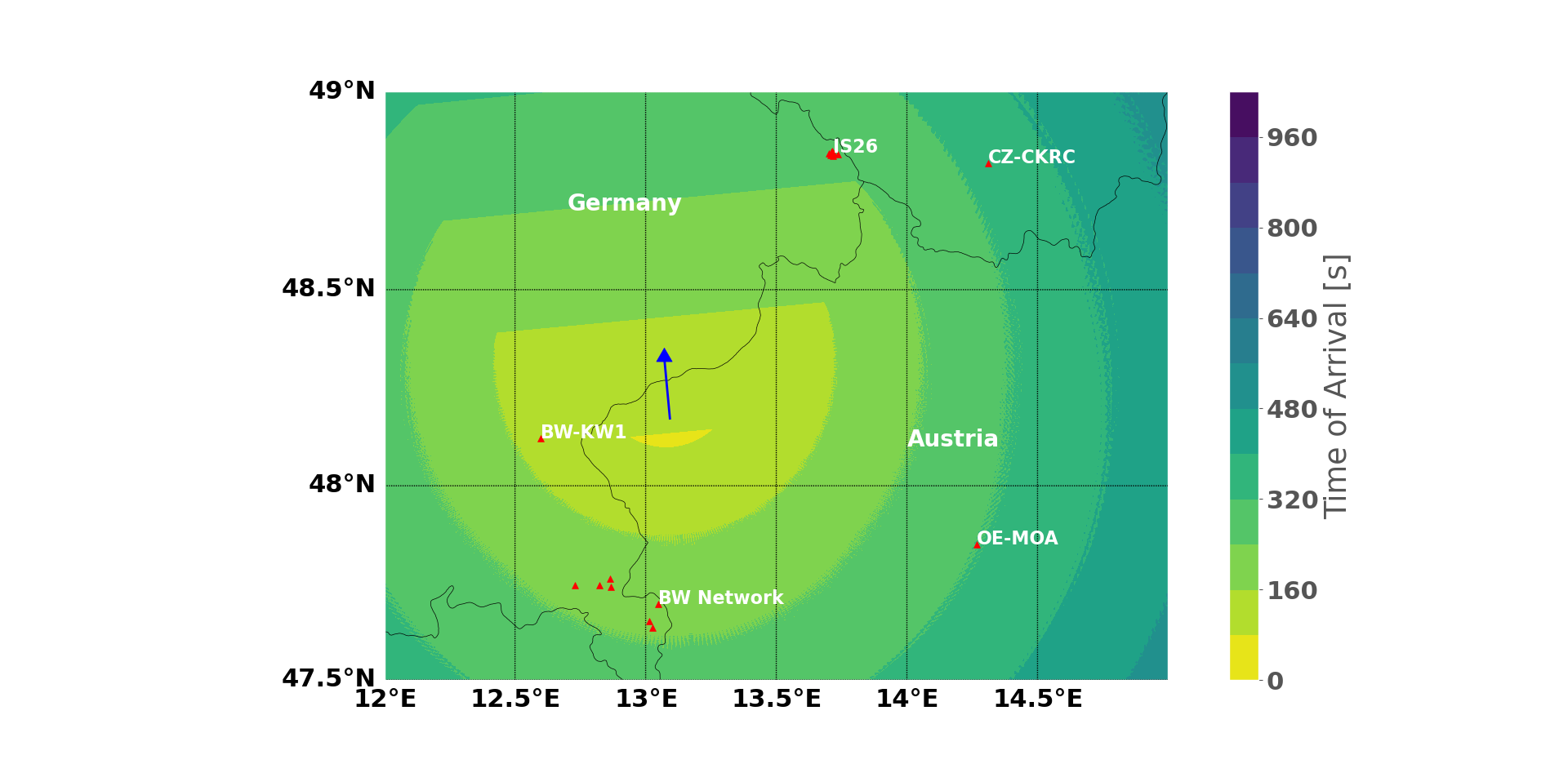}
        \caption{\corrbb{The ground footprint where the ballistic shock for an isotropic atmosphere emitted from the Stubenberg fireball is predicted to be detectable, assuming acoustic emission within a 25 degree opening angle of the velocity vector. Note that the ground area in front of the fireball is not predicted to detect the ballistic arrivals if the 25 degree tolerance is not applied. See Table \ref{tab:StubenbergTrajectory} for the complete Stubenberg trajectory, shown here as a blue arrow in the direction of travel taken from a height of 50 km to 17 km above the ground. \corrb{In our case study presented later, we use the IS26 infrasound array as our ground receiver for amplitude measurements.}}}
        \label{ball_no_wind}

\end{figure}

\begin{figure}[]

         \centering
         \includegraphics[width=\linewidth]{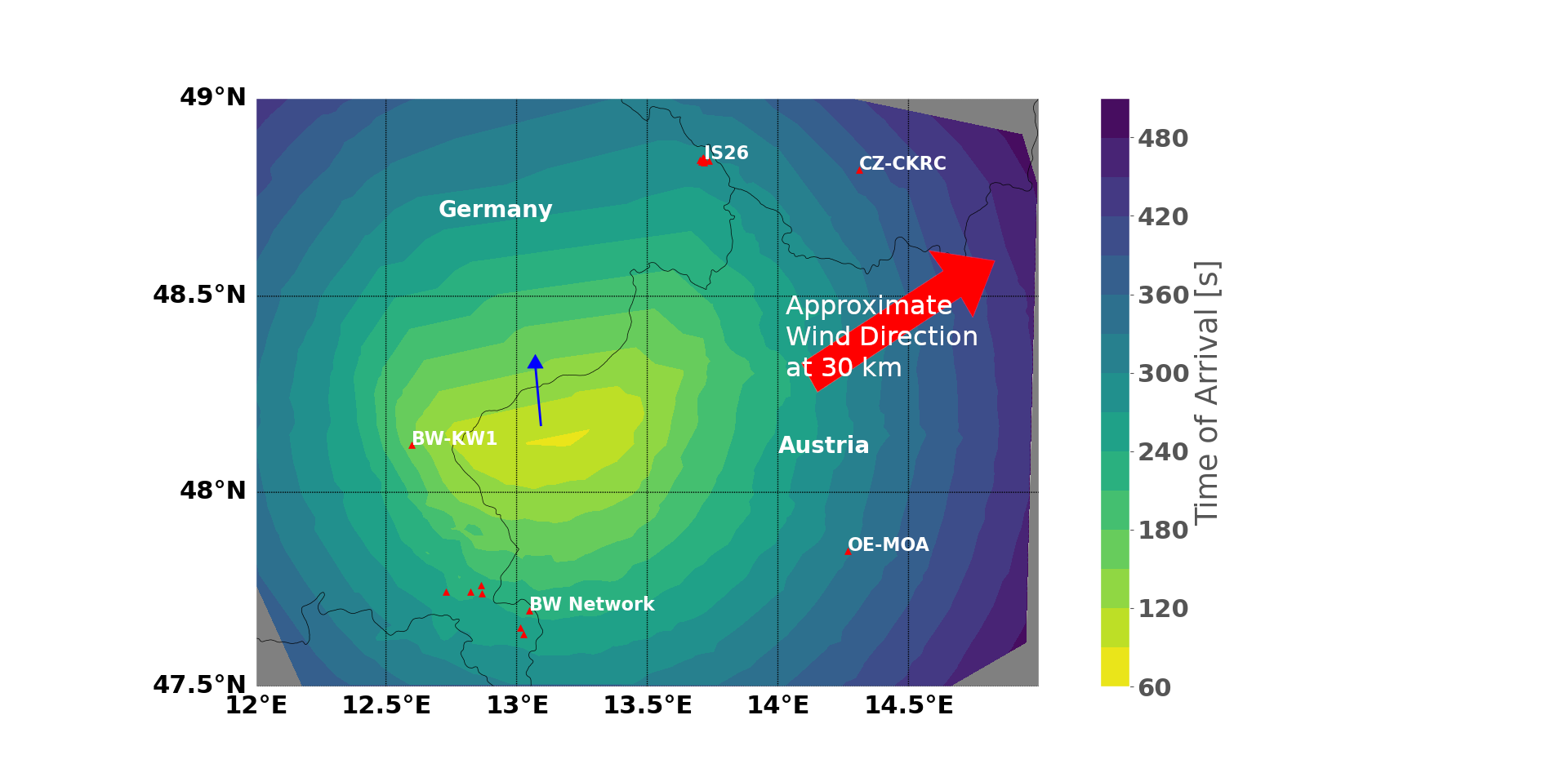}
        \caption{\corrbb{The same as Figure \ref{ball_no_wind}, except that arrival times are calculated using the model atmosphere rather than an isotropic atmosphere. The red arrow shows the approximate wind direction at a height of 30 km.}}
        \label{ball_wind}

\end{figure}
\end{subfigures}

\begin{subfigures}

\begin{figure}[]
  \centering

         \centering
         \includegraphics[width=\linewidth]{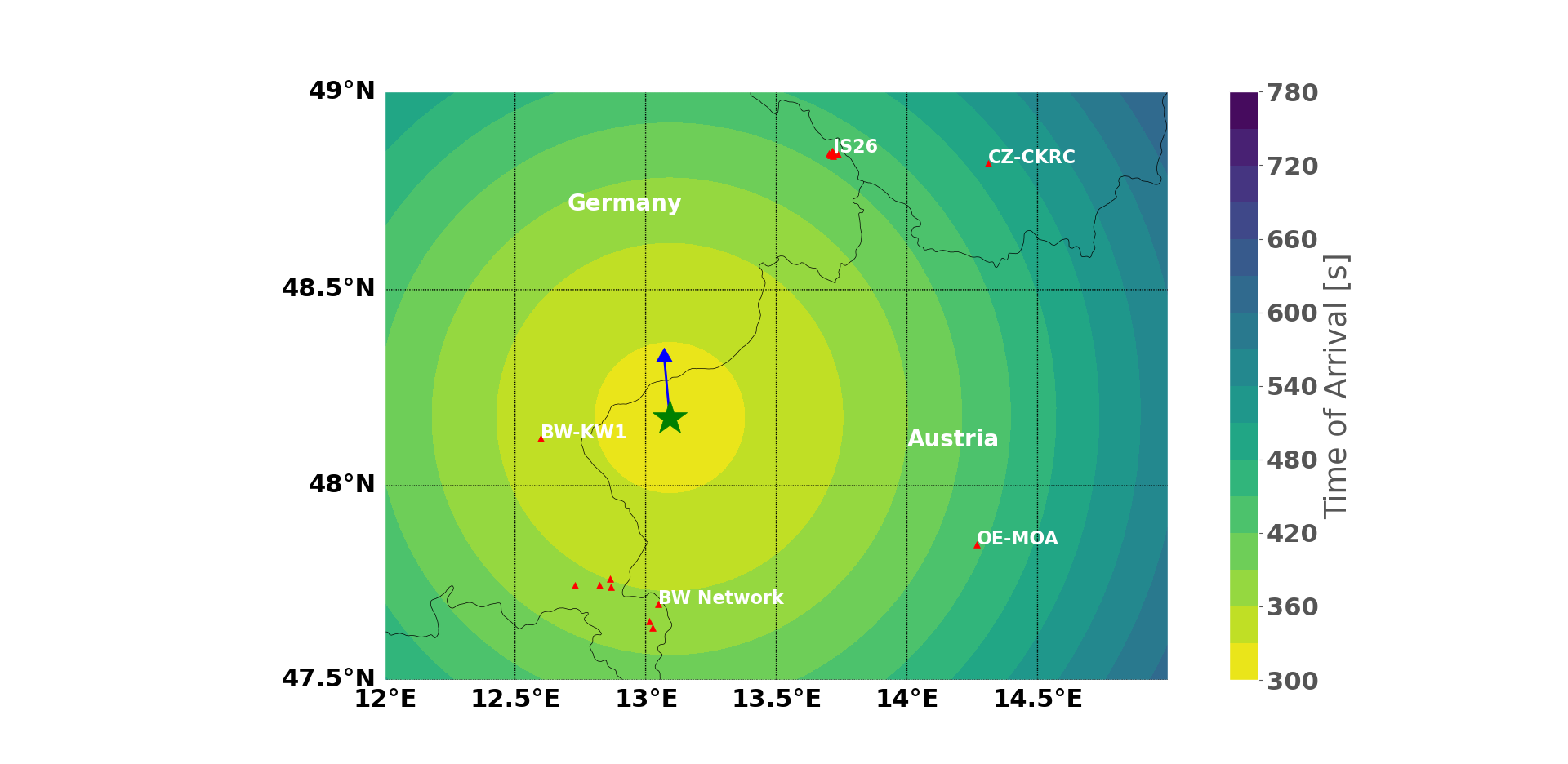}
         \caption{\corrbb{The ground footprint where rays from a fragmentation shock (shown by green star) are detectable for the Stubenberg fireball (shown in blue) in an isotropic atmosphere. Here an example fragmentation point is inserted at height of 50 km, at $48.17328^{\circ}$N, $13.09262^{\circ}$E. }}
        \label{frag_no_wind}

      \end{figure}

    \begin{figure}[]

         \centering
         \includegraphics[width=\linewidth]{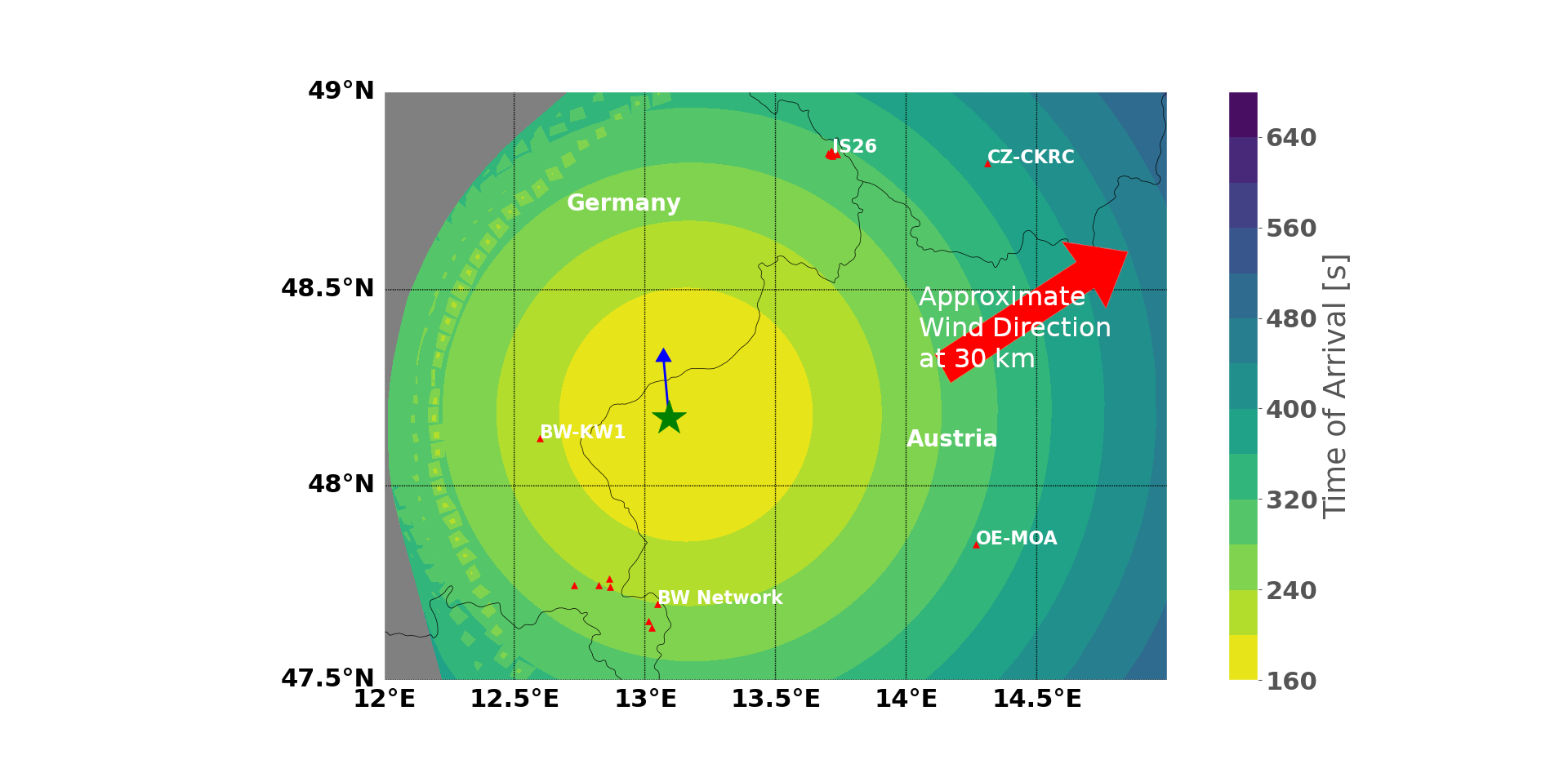}
         \caption{The same as Figure \ref{frag_no_wind} but in this case using a non-isothermal model atmosphere with winds. The red arrow shows the approximate wind direction at a height of 30 km.}
        \label{frag_wind}

\end{figure}
\end{subfigures}
The adiabatic speed of sound in a stationary medium depends on the temperature, as:

\begin{equation} \label{eq:speedofsound}
c_s = \sqrt{\frac{\gamma R T}{M_0}} \,,
\end{equation}
where for the Earth's atmosphere, below 90 km height, $\gamma = 1.40$, $M_0 = 28.9644$ g/mol, $R$ is the ideal gas constant ($R = 8.31 J K^{-1} $; The US Standard Atmosphere, 1976), and $T$ is the temperature in the medium in Kelvin \citep{Edwards2003, Revelle1974}.

The wind speed and magnitude also affects the propagation of fireball acoustics. For a given wind vector $\vec{w}$, the effective speed of sound $c_{eff}$ is:

\begin{equation} \label{eq:winds}
    c_{eff} = c_s + \hat{n} \cdot \vec{w} \,,
\end{equation}

\noindent where $c_s$ is the ambient speed of sound, and $\hat{n}$ is the wavefront normal \citep{Silber2019}. Combining Equations \ref{eq:speedofsound} and \ref{eq:winds}, we can model the speed and direction of the acoustic waves in the atmosphere. As the horizontal wind components are much larger than the small vertical winds, we ignore the latter. Therefore, the wind vector can be written as:
\begin{equation}
    \vec{w} = \left[ \begin{smallmatrix} u, v, 0 \end{smallmatrix} \right] \,,
\end{equation}
where $u$ and $v$ are the west to east, and south to north components of the wind vectors respectively. The components are converted into magnitude and direction as follows:
\begin{equation} \label{eq:mag}
    \left\lVert \vec{w} \right\rVert = \sqrt{u^2 + v^2} \,,
\end{equation}

\begin{equation} \label{eq:dir}
    \omega = \arctan{\frac{u}{v}} \,,
\end{equation}
\noindent where $\vec{w}$ is the wind vector of a specific layer, pointing in the direction the wind is blowing to, $u$ and $v$ are the speed of the Eastward and Northern components of the wind respectively. Here $\omega$ is the azimuthal angle of the wind vector, from the North, increasing to the East.
Equation \ref{eq:winds} gives the effective speed of the acoustic wave in any given layer of the atmosphere. \par

The modification to the fireball acoustic footprint at the ground under the action of the winds and \corrb{measured } atmospheric temperature structure is shown in Figures \ref{ball_wind} and \ref{frag_wind}. The effects of the atmosphere and its associated uncertainty impact the accuracy of the fireball trajectory and fragmentation hypocenters. The ground footprints also highlight regions where ground stations might detect fragmentation acoustics alone or both kinds of fireball sounds. \par
\corrb{For our modelling we use atmospheric data from the European Centre for Medium-Range  Weather Forecasts (ECMWF) \citep{Dee2011}, which supplies interpolated atmospheric data, including wind and temperature data, around the world. More details of the implementation of the atmospheric data within the program are described later.}

\subsection{Bolide Energy Estimation from regional Acoustic Measurements} \label{Shockwaves}

Meteoroids entering the atmosphere deposit their kinetic energy through drag interaction with atmospheric molecules. The shockwave thus produced begins as a strong (highly non-linear) shock \citep{Sakurai1964} which quickly decays to a weak-shock, and then transitions to a linear wave \citep{Revelle1976, Edwards2008, Silber2019}.\par

\corrb{Estimation of fireball source energy is complicated and poorly constrained by uncertainties in model interpolated atmospheric variables (including wind and temperature), which accumulate with range. }  The usual approach in such situations is to appeal to empirical relations between known explosive sources in the atmosphere and the resulting observed amplitudes (see \citealp{Silber2019} for a complete review). The source details of the fireball shock production are effectively scrubbed out at large ranges by the atmosphere \corrb{ as attenuation increases strongly at high frequencies \citep{Sutherland2004}. This effect tends to leave only the primary ``fingerprint" of the dominant fireball signal period (which for large bolides are at progressively longer periods, see equation \ref{af}, } and which is also less affected by propagation effects than is amplitude). This period can be roughly related to source energy using experimental relations. In particular, different fragmentation points or distinct arrivals from ballistic or fragmentation sources may merge at long ranges and thus become masked by propagation effects.

In contrast, at short, regional ranges ($<$200 km), it is often possible to distinguish discrete fragmentation points and isolate ballistic arrivals from fragmentation in acoustic records if some independent information is available concerning the fireball trajectory \citep[cf.][]{Borovicka2003b}. As a result, it also becomes possible to apply either analytic or numerical models to relate the period of direct acoustic arrivals at ground stations to properties of the airborne shock sources. 

\subsubsection{Acoustic energy estimate from Ballistic/Cylindrical shock} 
For the analytic approach, consider that the energy per unit path length, $E$, of a cylindrical shock is defined \citep{Edwards2008, Revelle1974, Plooster1970} as:
\begin{equation} \label{blastradii}
    E = R_0^2 P \,,
\end{equation}
where $R_0$ is the blast radius and $P$ the ambient atmospheric pressure in the portion of the trail of interest. $R_0$ is  defined as the radius of a volume of atmosphere that would be created if all of the explosion energy (resulting from drag) does pressure work to move the atmosphere from the centre of the trail outwards. The speed of the fireball produced shockwave goes from the order of 10 times the speed of sound, quickly down to the speed of sound within approximately one blast radius \citep{Sakurai1964}. The blast radius, also referred to as the relaxation radius, \citep{Few1969, Silber2019} is on the order of $\sim10^1 - 10^2$ m for most meteors of interest. This is several orders of magnitude less than the range rays typically travel to reach a station, of $\sim10^4 - 10^5$ m.\par
Assuming no ablation or fragmentation, this energy deposition can also be expressed in terms of the hypersonic drag law ($E = \frac{1}{2} \rho_a v^2 C_D A$), where $\rho_a$ is the atmospheric mass density, $v$ the meteoroid velocity, $C_D$ the aerodynamic drag and A the cross-sectional area of the body. Following the analysis of \cite{Revelle1974}, the cylindrical (or equivalently ballistic) blast radii with this simplification can be expressed as:
\begin{equation} \label{blastradiiD}
    R_0 = k d_m M \,,
\end{equation}
where $k$ is some constant of order unity, which depends on which definition of the blast radii is used. Using the definition given in Equation \ref{blastradii}, the value of $k$ is approximately 0.742 \citep{Revelle1976} ($k = \sqrt{\frac{\pi}{8} C_D \gamma}; \gamma = 1.4; C_D = 1$). Here $d_m$ is the entry diameter of the meteoroid, $\gamma$ is the ratio of specific heats of air and $M$ is the Mach number of the meteoroid, given by the ratio of the velocity of the meteor to the speed of sound:
\begin{equation}
    M = \frac{v}{c_{s}} \,,
\end{equation}
where $c_s$ is the speed of sound in air at the height of shock production. \par
Assuming a spherical meteoroid, ($d_m = \left( \frac{6 m}{\pi \rho} \right)^{1/3}$), the kinetic energy, $K$, of the meteor can be expressed in terms of Equation \ref{blastradiiD} as:
\begin{equation} \label{kinetic energy}
    K = \frac{1}{2} \left( \frac{\pi}{6 k^3} \rho \frac{R_0^3}{M^3} \right) v^2 \,,
\end{equation}
where $\rho$ is the density of the meteor.\par
This cylindrical-line source energy approach has been well used and explored in the literature \citep[cf.][]{Revelle1976, Edwards2009, Ens2012}. For direct arrivals, in particular, the fundamental period of the ballistic infrasound arrival correlates well to the optically estimated energy deposition for cm-sized meteoroids \citep{SilberBrown2015}. At larger sizes, the weak-shock period approach has been more difficult to validate, as short range infrasonic detection of well characterized meter-sized impactors are very rare and seismic detection of impacts rarer still. \corrb{Only the meter-sized Carancas meteorite fall, which produced a crater, has had its impact with the ground detected seismically \citep{Tancredi2009}}. At least one decimeter-sized meteorite-producing fireball has had short range infrasound detected and the energy estimate from the weak-shock cylindrical period approach agreed to within a factor of two of other independent techniques \citep{Brown2011}.\par
In contrast to acoustic energy estimates from the ballistic shock, no methodology has been proposed or applied to estimate released energy from regional acoustic measurements associated with fireball fragmentation. We develop a technique for estimating energy released at the point of a fireball fragmentation from observed infrasonic amplitude of direct acoustic arrivals in the next section.

\subsubsection{Acoustic energy estimate from Fragmentation shock}
Provided the energy deposition remains relatively constant, the blast radius changes slowly along the path of the meteoroid and the shock geometry is well approximated as a cylinder. When fragmentation occurs, the energy deposition changes rapidly, usually peaking over a small height interval and depositing a large fraction of the total available kinetic energy \citep{Wheeler2018}. While Equation \ref{blastradii} remains valid in this case, the strongly varying $E$ results in a strongly varying $R_0$ over a short segment of the trail, producing a quasi-spherical shock geometry.\par

The over-pressure measured at the ground associated with such a fragmentation event may be related to the energy released along the fireball path assuming the fragmentation episode is an idealized spherical explosive source. In this case the relaxation radius (or blast radius) is simply \citep{Few1969, Sakurai1964, Revelle1974, Jones1968, Tsikulin1970}:
\begin{equation}
     R_0 \propto \left (\frac{E_t}{P} \right)^{1/3}
 \end{equation}
where E$_t$ is the total energy of the spherical explosion and $P$, the ambient pressure at the explosion altitude. The coefficient of proportionality is dependant on the respective author's definition. \par


To estimate energy of a fireball fragmentation episode, we appeal to the results of \cite{KinneyGraham1985}. For a chemical explosion, they show that the overpressure from a spherical explosion in the atmosphere is given empirically as:
\begin{equation} \label{KG85}
    \frac{\Delta p}{P_S} = \frac{808 \left[1 + \left(\frac{Z}{4.5}\right)^{2}\right]}{\sqrt{1 + \left(\frac{Z}{0.048}\right)^{2}}\sqrt{1 + \left(\frac{Z}{0.32}\right)^{2}}\sqrt{1 + \left(\frac{Z}{1.35}\right)^{2}}} = f(Z) \,,
\end{equation}
 where $Z$ is the scaled range, $\Delta p$ is the measured overpressure (at an infrasound station at the ground), and $P_S$ is the ambient atmospheric pressure measured at the station. We will refer to Equation \ref{KG85} as the KG85 model. 
 
 Scaled range relates characteristics of a given blast to a standard, 1 kiloton of TNT (Trinitrotoluene) NE (Nuclear Explosives), explosion ($4.2 \times 10^{12}$ J), which we term the ``reference yield". A 1 kt TNT NE explosion is roughly equivalent to a 0.5 kt TNT HE (Chemical High-Explosives) explosion \citep{Reed1972a}. For example, a spherical explosion of yield $W$ at a range $R$ away from a detector would measure the same overpressure as an explosion of yield $1000W$ at a range of $10R$. \cite{KinneyGraham1985} define scaled range, $Z$, for a spherical explosion as:
\begin{equation} \label{scaleddistance}
    Z = f_d \frac{R}{(W/W_0)^{1/3}} \,,
\end{equation}
where $R$ is the range in metres, $W$ is the yield in Joules, $W_0$ is the reference yield, and $f_d$ is a transmission factor. This latter term takes into account the varying atmospheric densities the acoustic wave traverses through the altitude change from the source ($h_2$) to the receiver ($h_1$), and is defined as:
\begin{equation} \label{transmission}
    f_d = \frac{1}{h_2 - h_1} \left(\frac{T_0}{P_0}\right)^{1/3}{\int^{h_2}_{h_1} \left(\frac{P(h)}{T(h)}\right)^{1/3} dh} \,,
\end{equation}
with $P$ and $T$ representing the pressure and temperature as a function of height, and $P_0$ and $T_0$ representing the pressure and temperature of a reference explosion, taken here as at the ground with the standard atmosphere and pressure values, where $P_0 = 101325$ Pa and $T_0 = 288 $K. 

\par
From Equation \ref{KG85}, as the range becomes large, the overpressure goes as the inverse of the range. As noted in \cite{KinneyGraham1985}, the sound wave is further affected by atmospheric attenuation. Thus, intrinsic atmospheric attenuation and geometric attenuation associated with refraction must be included to properly estimate source energy. \cite{Latunde2013} expanding on the work of \citet{Reed1972, Reed1972a}, incorporates both attenuation and geometric factors to model a pressure pulse propagating through the atmosphere. The attenuation factor for such a pressure wave is given as:
\begin{equation}
    af = \frac{\Delta p_{af, nonideal}}{\Delta p_{af, ideal}} \,,
\end{equation}
where $\Delta p_{af, nonideal}$ is the attenuated amplitude of the pressure wave, and $\Delta p_{af, ideal}$ is the amplitude of the pressure wave in an ideal, isotropic atmosphere. We approximate the path the acoustic path takes as a series of straight lines. \citet{Latunde2013} models the attenuation between atmospheric layers $n+1$ and $n$ as:
\begin{equation}
    af_n = \frac{\Delta p_{af, n+1}}{\Delta p_{af, n}} = \exp{\left(-\frac{k \nu^2}{b P_0 \sin{\theta_{n}}}\left[e^{b h_{n+1}} - e^{b h_{n}}\right]\right)} \,.
\end{equation}

The total atmospheric attenuation factor, $af$, is therefore found as:
\begin{align} \label{af}
    af &= \frac{\Delta p_{af, n}}{\Delta p_{af, 0}} \,, \nonumber \\
       &= \prod_{i=0}^{n-1} \frac{\Delta p_{af, i+1}}{\Delta p_{af, i}} \,, \nonumber \\
       &=  \exp{\left( \sum_{i=0}^{n-1}  -\frac{k \nu^2}{b P_0 \sin{\theta_{i}}}\left[e^{b h_{i+1}} - e^{b h_{i}}\right]\right)} \,,
\end{align}
and the refraction correction, $rf$, is given as \citep{Latunde2013}:

\begin{equation} \label{rf}
    rf = \frac{\Delta p_{rf, n}}{\Delta p_{rf, 0}} = \sqrt{\frac{\frac{d \theta_0}{dr}_{nonideal}}{\frac{d \theta_0}{dr}_{ideal}}} \,,
\end{equation}

\noindent where

\begin{description}
\item  $\Delta p_{0}$ is the overpressure which would be measured at the station, if the atmosphere did not affect the overpressure at all. Equivalently, this is the overpressure measured at the source.
\item  $\Delta p_{n}$ is the overpressure measured at the station, after travelling through n-layers of atmosphere.
\item $k$ is a constant, typically given as $2.0 \times 10^4$ kg m$^{-2}$ \citep{Reed1972, Reed1972a}; other works adopt values within a factor of two of this value. \cite{MorseIngard1968}, for example, state both $1.371 \times 10^4$ kg m$^{-2}$ and $3.047 \times 10^4$ kg m$^{-2}$ as possible values for $k$.
\item $\nu$ is the fundamental frequency of the acoustic wave.
\item $b$ is the inverse of the scale height of the atmosphere, which we adopt as $1.19 \times 10^4$ m$^{-1}$ \corrb{(The US Standard Atmosphere, 1976)}. Values for $b$ change as a factor of $\frac{R_e}{R_e + h}$, where $R_e$ is the radius of the Earth, and $h$ is the height the scale height is evaluated at. This study uses source-receiver heights generally below 50 km, and $b$ changes by less than $1\%$ over this distance. Physically: $b = \frac{Mg}{RT}$, where $R$ is the ideal gas constant, $T$ is the mean atmospheric temperature, $g$ is the acceleration due to gravity, and $M$ is the mean molar mass of atmospheric molecules.
\item $P_0$ is the standard pressure.
\item $\theta_i$ is the angle of depression of the direct ray in a given atmospheric layer.
\item $h_i$ is the height of atmospheric layer $i$.
\item $\frac{d \theta_0}{dr}_{nonideal}$ is the ratio of the ray takeoff azimuths from the source to the ground area where a ray-traced solution ends, in a realistic atmosphere.
\item $\frac{d \theta_0}{dr}_{ideal}$ is the ratio of the ray takeoff azimuths from the source to the ground area where a ray-traced solution ends, in an isotropic atmosphere.
\end{description}
Equation \ref{KG85} assumes that the overpressure measured is $\Delta p_0$, without accounting for attenuation. Therefore, using Equations \ref{af} and \ref{rf} we obtain:
\begin{align} \label{pressure_eq}
    \frac{\Delta p_0}{P_S} = f(Z) = \frac{\Delta p_n}{(P_S \times {af} \times {rf})} \,, \nonumber \\
    {\Delta p_n} = f(Z) ( P_S \times af \times rf ) \,.
\end{align}

    
Using Sach's scaling as outlined in \cite{Reed1972, Reed1972a}:
\begin{equation} \label{sach}
    \nu^2 = \frac{1}{4 J_0^2} \left( \frac{W_0 P}{W P_0} \right)^{2/3} \left( \frac{c}{c_0} \right)^2 \,,
\end{equation}
and making use of Equation \ref{scaleddistance}, we can write overpressure at the ground in terms of energy released at a single fireball fragmentation point as:
\begin{equation} \label{g}
\begin{split}
    \Delta p_n &= f\left(\frac{f_d R}{\left(W/W_0\right)^{1/3}}\right) ( P_S \times af(W, P, h, R) \times rf ) \\&= g(W, P, h, R) \,,
\end{split}
\end{equation}
where:
\begin{equation}
    af(W, P, h, R) = \exp \left( {W^{-2/3} F(P, h, R)} \right)
\end{equation}
and
\begin{multline}
    F(P, h, R) = \\
    \frac{k}{4 b J_0^2} \left( \frac{W_0 P}{P_0} \right)^{2/3} \left( \frac{c}{c_0} \right)^2  \sum_{i=0}^{n-1} {\left( -\frac{\left[e^{b h_{i+1}} - e^{b h_{i}}\right]}{P_{0} \sin{\theta_{i}}}\right)} \,,
\end{multline}
\noindent and where

\begin{description}
\item $J_0$ is the positive phase duration of the reference explosion, taken here to be $0.375$ s \citep{Reed1972, Reed1972a}.
\item $c_0$ is the speed of sound of the reference explosion, taken here to be 347 m s$^{-1}$ \citep{Reed1972, Reed1972a}.
\item $c$ is the speed of sound at the height of the explosion.
\end{description}

For a given fragmentation event occurring at a known height, pressure, and range from the sensor where overpressure is measured at the ground, $g(W, P, h, R) = g(W)$ and the yield can be found via:
\begin{equation} \label{g_inv}
    W = g^{-1}(\Delta p_n).
\end{equation}

Function $g$, as defined in Equation \ref{g}, is difficult to invert analytically in terms of yield, so we solve it numerically.\par

For both the cylindrical and spherical shock geometries, the non-linear shock inside $R_0$ (where $\frac {\Delta P}{P}$ $>$ 1) is followed by a slower amplitude decay in the weak-shock regime \citep{Jones1968, Plooster1970, Revelle1974} which extends outward many tens to hundreds of blast radii until the shock approaches a linear acoustic wave. From Equations \ref{g} and \ref{g_inv} we now have a technique to estimate fragmentation energy purely from acoustic amplitudes measured in the linear regime, provided the signal is known to be from a fragmentation point. The complementary analytic expressions for amplitude and period evolution of the shock for energy deposition in cylindrical geometry is well summarized in \cite{Revelle1974, Edwards2009, Silber2019} and is not repeated here. \par
We note that the most robust means of estimating the meteor-associated shock amplitude and period from either a cylindrical or fragmentation-type source as a function of release height and range is to employ a full Computational Fluid Dynamics (CFD) approach, sourcing the energy to initiate the numerical simulation and following the shock to the ground observing point \citep[cf.][]{Nemec2017}. \corrb{CFD would include non-linear effects to ray paths, such as scattering and dispersion of acoustic waves. It is unlikely to affect travel times significantly, but CFD would permit synthetic waveforms to be computed and a more robust comparison between observed wavetrains and the source function to potentially be studied, including frequency-dependent effects on amplitudes. This might improve source energy estimates. } Presently we restrict our fragmentation energy estimate to this simple semi-empirical analytic approach, but hope to eventually compare to a full CFD solution. \par
\subsection{Meteoroid speeds estimated from acoustic arrivals}

\begin{figure}[ht!]
\vspace*{2mm}
\begin{center}
\includegraphics[width=\linewidth]{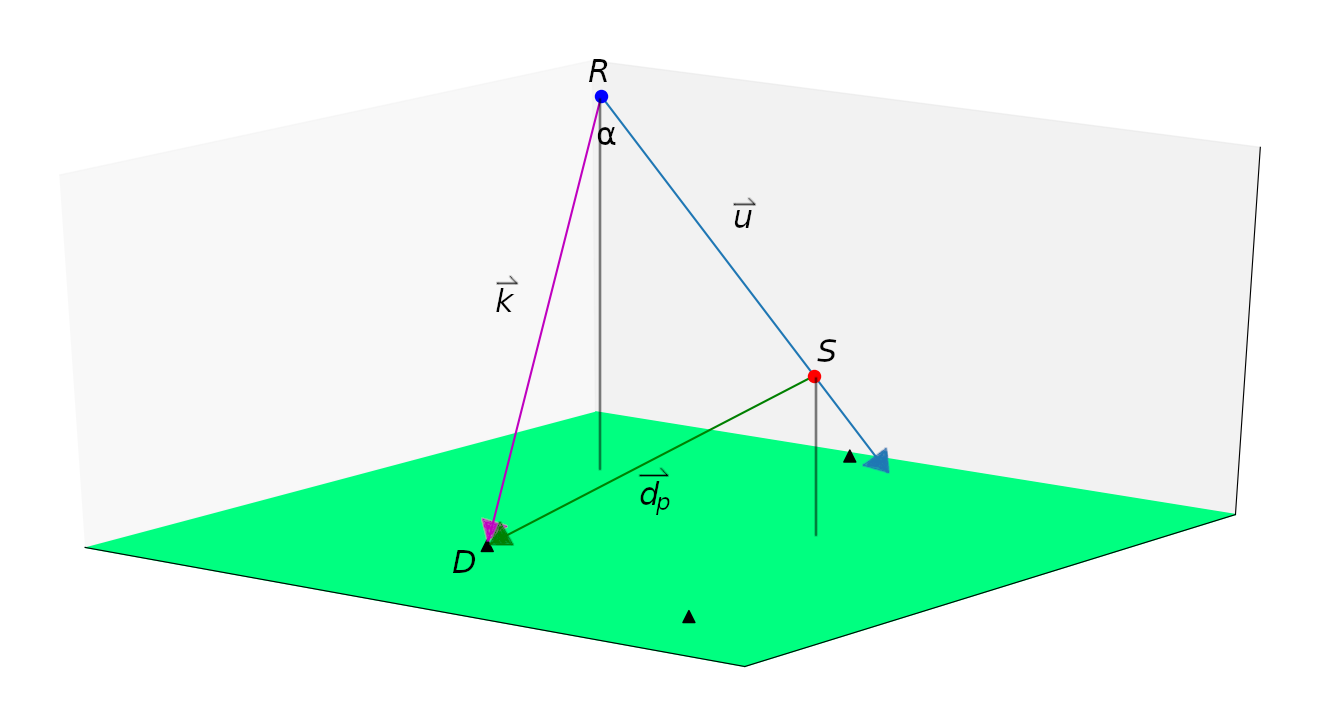}
\caption{A diagram showing the vectors used in calculating the time of an arrival relative to a reference point $R$. $S$ is the wave release point of the acoustic wave, and $D$ is the seismic or infrasound station.}
\label{ballisticdiagram}
\end{center}
\end{figure}

The timing of an acoustic signal from a fireball at a seismic or infrasound station is the addition of two times, as shown in Figure \ref{ballisticdiagram}: $t_{traj}$, the time the fireball takes to travel from a common reference point, $R$, along its trajectory to a wave release point, $S$, and $t_{ray}$, the acoustic travel time from the wave release point to the station, $D$. In an isotropic atmosphere, it is trivial to show that the total relative time, $t_{total}$ for the acoustic signal to arrive at the station is:

\begin{align} \label{wrp}
    t_{total} & = t_{traj} + t_{ray} \,, \nonumber \\
              & = \frac{\left\lVert{\vec{k}}\right\rVert \cos{\alpha}}{v} + \frac{\left\lVert{\vec{k}}\right\rVert \sin{\alpha}}{c} \,,\nonumber \\
              & = \frac{\vec{k} \cdot \hat{u}}{v} + \frac{\left\lVert{\vec{k} \times \hat{u}}\right\rVert}{c} \,,
\end{align}




\noindent where $\vec{k} = D - R$ is the vector from the reference point to the station, $\hat{u}$ is the unit vector of the trajectory, $v$ is the velocity of the fireball, assumed to be constant, $c$ is the speed of sound, assumed to be constant, and $\alpha$ is the angle between $\vec{k}$ and $\vec{u}$. 
With a known wave release point, $S$, Equation \ref{wrp} becomes:
\begin{equation} \label{wrp_simple}
    t_{total} = \frac{\left\lVert{\vec{u}}\right\rVert}{v} + \frac{\left\lVert{\vec{d_p}}\right\rVert}{c} \,,
\end{equation}
where $\left\lVert{\vec{u}}\right\rVert$ is the length along the trajectory to $S$, and $\left\lVert{\vec{d_p}}\right\rVert$ is the distance from $S$ to the station. \par
By Equation \ref{wrp_simple}, changing the velocity of the fireball will change $t_{traj}$ and therefore $t_{total}$, but will not change $t_{ray}$. Therefore:

\begin{equation}\label{wrp_simplified}
    t_{total}(v) = \frac{\left\lVert{\vec{u}}\right\rVert}{v} + t_{ray} \,,
\end{equation}
where $t_{ray}$ is a constant time for a given trajectory, station, and speed of sound. \par
For a theoretical fireball travelling infinitely fast, $t_{total}(v \rightarrow \infty) = t_{ray}$. For a fireball of finite velocity, we define $\Delta t$ as the offset in total time from the infinitely fast fireball, such that:

\begin{align} \label{theo_eq}
    \Delta t(v) &= t_{total}(v) - t_{total}(v \rightarrow \infty) \,, \nonumber \\
               &=\frac{\left\lVert{\vec{u}}\right\rVert}{v}\,,\nonumber\\
                &= \frac{\Delta h}{v \cos \theta} \,,
\end{align}


where $\Delta h$ is the difference in heights between the reference position and the wave release point, and $\theta$ is the zenith angle of the fireball. The offset provides a metric to see how resolvable a specific velocity is at a station relative to the time pick precision. It determines how much later the acoustic signal is expected at a station due to the finite velocity of the fireball. If two fireball velocities have similar offsets, then it will be difficult to distinguish between them on a waveform. However, if two velocities have offsets that differ on the order of the variation expected due to atmospheric uncertainties alone and differ by an amount greater than the pick precision, then the velocity may be resolvable . \par
Therefore, the offset in time of a fireball is proportional to the change in height between the reference and the wave release point, its velocity, and its zenith angle. From Equation \ref{theo_eq}, it is easier to resolve a fireball velocity if the wave release point occurs deeper in the atmosphere or if wave release points for various stations are far apart from each other, the velocity of the fireball is relatively low, and for grazing trajectories. \par
In a realistic atmosphere, Equation \ref{wrp_simple} becomes more complicated, as the path between $S$ and $D$ becomes curved, and the effective speed of sound, $c$, changes in 3-D space. However, for a given trajectory, station, and a specific atmosphere, the travel time of the acoustic ray will be constant, therefore we may use Equation \ref{wrp_simplified} in general, keeping the assumption that the velocity of the fireball is constant. \par
As before, we assume the atmosphere model is uncertain so both the travel time of the ray and the wave release point will have some uncertainty. Therefore, Equation \ref{wrp_simplified} becomes:
\begin{equation*}
    t_{total}(v) \pm \delta t_{total}(v) = \frac{(\left\lVert{\vec{u}}\right\rVert \pm \delta \left\lVert{\vec{u}}\right\rVert)}{v} + (t_{ray} \pm \delta t_{ray}) \,,
\end{equation*}
and the definition of offset, $\Delta t (v)$ remains as it is in Equation \ref{theo_eq}. Thus in some very limited instances with good station geometry and for shallow entry trajectories we might expect to be able to constrain fireball speeds with modest precision.

\section{Methodology/Implementation within the \texttt{BAM} code}

\subsection{Overview}

\texttt{BAM} is written in Python and is a GUI-based signal waveform analysis, identification, and ray-tracing package. Its purpose is to allow easy identification of common infrasonic or seismo-acoustically coupled signals originating directly from fireballs. It is designed for direct arrivals and not ducted signals; hence it is limited to ranges of order $\sim$200 km or less for fireballs. \corrb{More details on the \texttt{BAM} software, including the repository and the user manual, are available in \ref{BAMdocs}.} \par
In most common situations, a fireball time and approximate location are known from other sources (eg. eyewitness visual, dashcam, optical cameras) and this provides a framework for locating and identifying associated fireball acoustic signals. Once signals are identified manually and tagged as either fragmentation or ballistic related, \corrb{\texttt{BAM} allows } for the automated geolocation of both fragmentation points and trajectory orientation. \par
Having the approximate fireball location identified, all available stations (mostly seismic, but some infrasound if available) within 150 km ground range are \corrb{used for analysis}. A full reference of seismic and infrasound networks used in this study is given in \ref{stationref}. 

\subsection{Atmosphere}
\corrb{This study explores the effect atmospheric model accuracy has on a fireball acoustic solution, } through atmospheric data from the European Centre for Medium-Range Weather Forecasts (ECMWF) (obtained from the Copernicus Climate Change Service\footnote{Copernicus Climate Change Service (C3S) (2017): ERA5: Fifth generation of ECMWF atmospheric reanalyses of the global climate. Copernicus Climate Change Service Climate Data Store (CDS). https://cds.climate.copernicus.eu/cdsapp\#!/home},  \citealp{Dee2011}). These specific sources were chosen because of their data assimilation methods, \corrb{which include both nominal atmospheric data and error estimates in the form of atmosphere ensemble members } \citep{Dee2011}. The observations and measurements from the model interpolates the atmospheric profiles continuously over land, rather than only where the measurements were taken. Atmospheric profiles, also referred to as atmospheric soundings, are a vertical measurement of various parameters, such as temperature or wind speed, as a function of pressure, or height. The data assimilation model provides sounding data within a spatial and temporal grid. 

The model is used to estimate the propagation of the acoustic waves from the fireball to seismic and infrasound stations located within approximately 200 km of the trajectory \citep{Brown2003}. ECMWF was chosen in particular due to its ensemble member calculations which provides atmospheric perturbation ranges that may be used in conjunction with the nominal atmospheric data to predict uncertainties.\par

The atmospheric profile is generated by the model at specific places and times from an ensemble of observations globally fit by the model. In reality there is some variance in the temperatures and wind not captured by the model. To simulate the real variance expected in the atmosphere and estimate the corresponding uncertainty in shock arrival time, a perturbation scheme is adopted through the ensemble variations provided by the ECMWF model \citep{Dee2011}. \par

With these realizations, the perturbation method embedded within \texttt{BAM} allows the magnitudes of the temperatures and wind components to vary within a physically reasonable range. This is then used to establish variations in estimated acoustic ray-tracing travel times and arrival azimuths from the fireball to a given station and hence provide a measure of the underlying solution uncertainty driven by uncertainties in the atmospheric model. \par

To calculate the uncertainty in the arrival times, each arrival using the nominal atmosphere is bracketed by timing using the perturbed atmosphere. For each solution, ballistic or fragmentation, the nominal solution is found by inverting the arrival times at each station, and the perturbed arrival times are then propagated through new trajectory solutions to find the uncertainty in the trajectory parameters or supracenter location. More details of this approach are provided in \ref{BAMdocs}. \par

\subsection{Fragmentation points : Supracenter}\label{list}
To locate fragmentation points from acoustic arrivals, \texttt{BAM} builds on the Supracenter module of \citet{Edwards2004}. Here we use the term ``supracenter" to refer to the four-dimensional position of the fragmentation point. If the location and time are unknown, a minimum of four stations is required to isolate a single fragmentation point and its timing \citep[e.g.][]{Tatum1999}. \corrb{More information about the supracenter inversion process can be found in \ref{BAMdocs} and in \cite{Edwards2003}.}\par

\corrb{The propagation from source, $S$, to detector, $D$, uses the Tau-P ray-tracing method \citep{Garces1998} found in \cite{Edwards2003}. Specific details on how \texttt{BAM} computes ray paths may be found in \ref{BAMdocs}. This method iteratively searches for the optimal launch zenith and azimuth angles for the ray, taking into account the atmosphere from $S$ to $D$. The Tau-P method uses a purely linear acoustics approximation, when in reality, non-linear effects, such as scattering, may allow rays otherwise trapped in ducts at higher altitudes to propagate to the ground. These false positives tend to overestimate the size of ray shadow zones within our models. To mitigate these effects, we increase the number of attempted ray-traces for each solution, with higher precision zenith and azimuth launch angles. In an attempt to correct for non-linear effects, we allow for both a horizontal and vertical tolerance for the ray to miss a given station to account for scattering effects.}

It is assumed that all propagation from $S$ to $D$ is purely linear acoustics. Near the meteor where the shock pressures are very high this assumption is invalid.  However, as discussed in Section \ref{Shockwaves}, the range the wave travels in this regime compared to in the linear regime is orders of magnitude smaller, and therefore will not have a large affect on the travel time, though it may affect the launch direction. 

\subsection{Acoustically-derived fireball Trajectory}
The Acoustic Trajectory module is designed to estimate the fireball trajectory given a set of at least six station ballistic arrival times, this being the required minimum for an estimate \citep[e.g.][]{Tatum1999} in the absence of any timing information (only five are required if the fireball time is known). The parameters describing the trajectory include the latitude and longitude of the geometric intersection of the assumed linear fireball path with the ground (termed the terminal ground point), the azimuth and zenith angle of the trajectory heading, and the time and velocity of the fireball. These parameters are shown in Figure \ref{ballistic_model}. 
All ray tracing is calculated using the Tau-P ray-tracing algorithm as in \texttt{SUPRACENTER}. For ballistic waves, the wave release point is the point along the trajectory where acoustic waves are released perpendicular to the trajectory vector (the specular point from a particular receiver). This point is simple to find geometrically; if there are no winds and temperature changes, the wave travels in a straight line from the trajectory to the station. However, the path becomes complex when the rays curve under the action of winds and temperature changes. To avoid this complication, various sample points along the trajectory have their ray paths calculated, and the point with the initial launch angle closest to, within a tolerance of $\sim 25^{\circ}$ \citep{BROWN2007}, to perpendicular to the trajectory is used as the wave release point.\par


\subsection{Forward modelling of Fragmentation Timing} \label{heightestimates}
\begin{figure}[ht!]
\vspace*{2mm}
\begin{center}
\includegraphics[width=\linewidth]{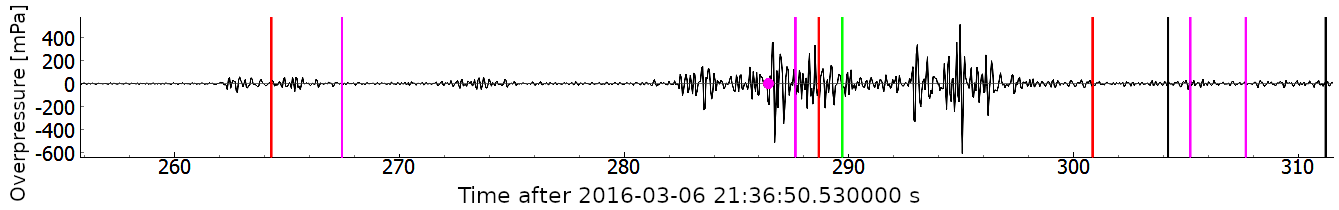}
\caption{An example of acoustic arrivals from potential fragmentation points at various heights along a known fireball trajectory (in this case the Stubenberg meteorite producing fireball) superimposed on waveform data at an infrasound station. The different coloured vertical lines represent arrivals from various source heights, (increasing in height from left to right). The magenta dot represents a user sample time pick, used in producing Figure \ref{examplestaff}.}
\label{arrivals}
\end{center}
\end{figure}

\begin{figure}[ht!]
\vspace*{2mm}
\begin{center}
\includegraphics[width=\linewidth]{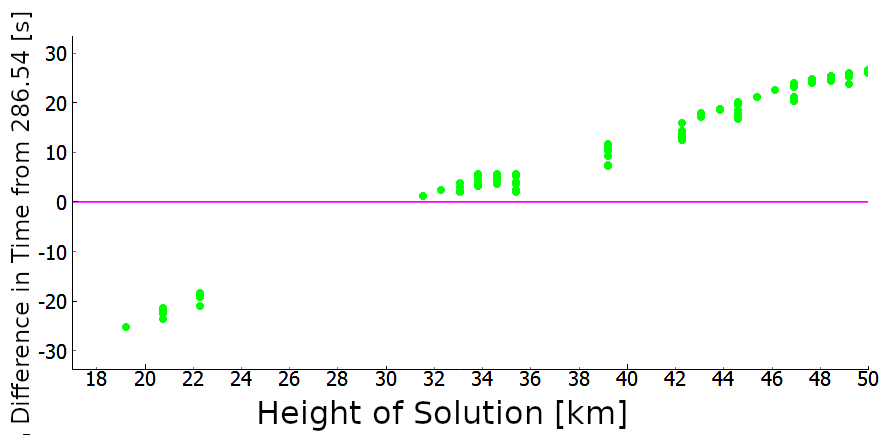}
\caption{A representation of timing of arrivals at one station from given heights along a known fireball trajectory. The magenta line represents the user-defined time shown as the magenta dot in Figure \ref{arrivals}. The green dots represent the range in arrival time for both nominal and perturbed arrivals from trial heights along the trajectory. In this example, based on the signal timing pick in Figure \ref{arrivals} the fragmentation point is most consistent with a height of approximately 31.5 km. This result agrees with optical measurements (See Figure \ref{lightcurve})}
\label{examplestaff}
\end{center}
\end{figure}

If a test fireball trajectory is known, Supracenter can be used to predict the time of travel for acoustic waves from many points of the trajectory and plot these at each station for manual inspection. Figure \ref{arrivals} shows an example of expected arrival times from varying heights along a fireball trajectory superimposed on a station waveform. This utility permits association of multiple coda in a signal waveform with the location and timing of fragmentation points determined by other techniques (e.g. optically measured). The user can apply this information to make a signal time pick (shown here as a magenta dot) and the relative arrivals of all other simulated fragmentation points are displayed (Figure \ref{examplestaff}), corrected for position along the fireball trajectory. This graph is useful for finding the probable source heights on the trajectory of different picks made along the waveform. The magenta line in Figure \ref{examplestaff} corresponds to the magenta dot on Figure \ref{arrivals}, while the green dots in Figure \ref{examplestaff} represent both the nominal and perturbed arrival times shown in Figure \ref{arrivals}.

\subsection{Energy estimation with \texttt{BAM}}
\subsubsection{Fragmentation Energy}

\begin{figure}[ht!]
\vspace*{2mm}
\begin{center}
\includegraphics[width=\linewidth]{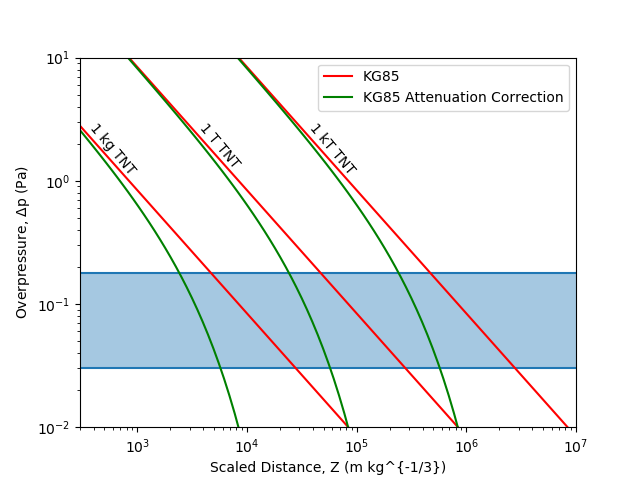}
\caption{The overpressure vs. scaled range of an example fragmentation from 32.7 km in height (Fragmentation 5 in the Stubenberg case study as described in Section \ref{sec:case}). The KG85 model (red lines) overestimates the pressure, and therefore underestimates the yield from an explosion as it does not include attenuation (shown by the green curves which are the ones implemeneted in \texttt{BAM}). Note that for a 1 kT explosion, scaled range is the actual range. The shaded region gives the range of pressures observed for the Stubenberg fireball.}
\label{figKG85}
\end{center}
\end{figure}

As shown in \cite{KinneyGraham1985} by Equation \ref{KG85}, the overpressure measured at a station can be expressed in terms of the scaled range. Figure \ref{figKG85} shows this model as implemented in \texttt{BAM} with an example attenuation correction for the Stubenberg fireball. The height of the fragmentation  and range of the propagation are required to generate these curves, as well as the atmospheric data along the acoustic path. The height of the fragmentation was interpolated from the waveform data, using the method described in Section \ref{heightestimates} and confirmed by the optical records. \par

The overpressure of a fragmentation-related acoustic signal is found from an infrasound record, by measuring the average of the positive and negative peaks of the associated bandpassed pressure waveform. The bandpass used and the number of fragmentation episodes assumed by the user affects the total energy. For the former, the bandpass is adjusted to be just above the dominant frequency of the fragmentation following a process for selecting bandpasses for fireball airwaves described in \citet{Ens2012}. We find that in pratice for most fireballs of interest, a bandpass of 0.1 - 2 Hz is appropriate.  \par


With the functional form of $g^{-1}(\Delta p_n)$ from Equation \ref{g} we therefore have $W$ as a function of $\Delta p_n$ for a given fragmentation location, station, and atmosphere. From the known fragmentation height and the observed overpressure, the discrete energy deposition of a fragmentation episode is measurable. 

\subsubsection{Ballistic Energy}
\label{sec:balen}
If ballistic arrivals are identified at one or more stations, measurement of the period and peak amplitude at those stations using the \texttt{BAM} \corrb{software } can be translated into a total energy estimate for the fireball. Here the assumption is made that the fireball ablates as a single body and the equivalent body diameter at the ballistic launch point can then be found. \par
The resulting energy deposition estimates per unit path length are made using Equation \ref{blastradii} and \ref{blastradiiD}. The blast radius is estimated  from the observed amplitude, period and known range to the fireball following the method of \cite{Revelle1976}. This procedure \citep{Edwards2009, Silber2019} produces estimates of the expected period and maximum amplitude at a station from a cylindrical shock with known $R_0$, under the assumption that the signal propagates as either a linear wave or as a weak-shock. By iteratively comparing the model predicted period and amplitude to the observed waveform at a station, a self-consistent blast radius is found. This can then be related directly to the meteoroid diameter and through the known fireball speed to the total kinetic energy of the body at that point.  

\section{Case Study}
\label{sec:case}
To demonstrate the methodology just described we apply \texttt{BAM} to a fireball where a meteorite was recovered and for which trajectory and energy estimates are available from optical records, namely the Stubenberg fireball.
\subsection{Overview}
\begin{figure*}[]
\begin{center}
\includegraphics[width=\linewidth]{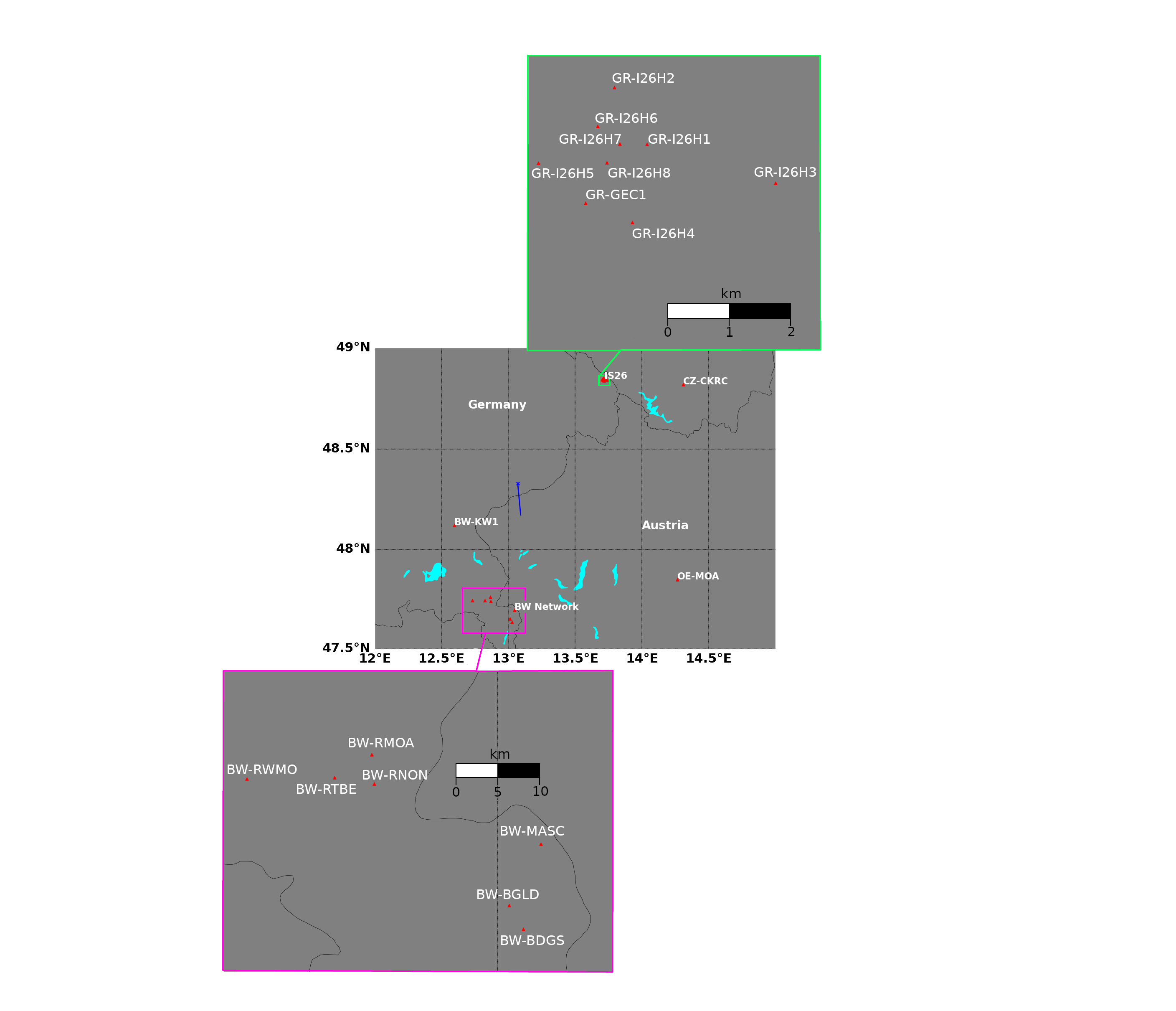}
\caption{Locations of nearby (acoustically accessible) stations of the Stubenberg fireball. The fireball ground track is shown in blue and the cross represents the geometric landing point in the centre figure. The local seismic station names referred to in the paper are shown in the lower blow up plot and the individual elements of the I26 infrasound array shown in the upper plot.}
\label{groundtrack}
\end{center}
\end{figure*}

\begin{figure*}[ht!]
\vspace*{2mm}
\begin{center}
\includegraphics[width=\linewidth]{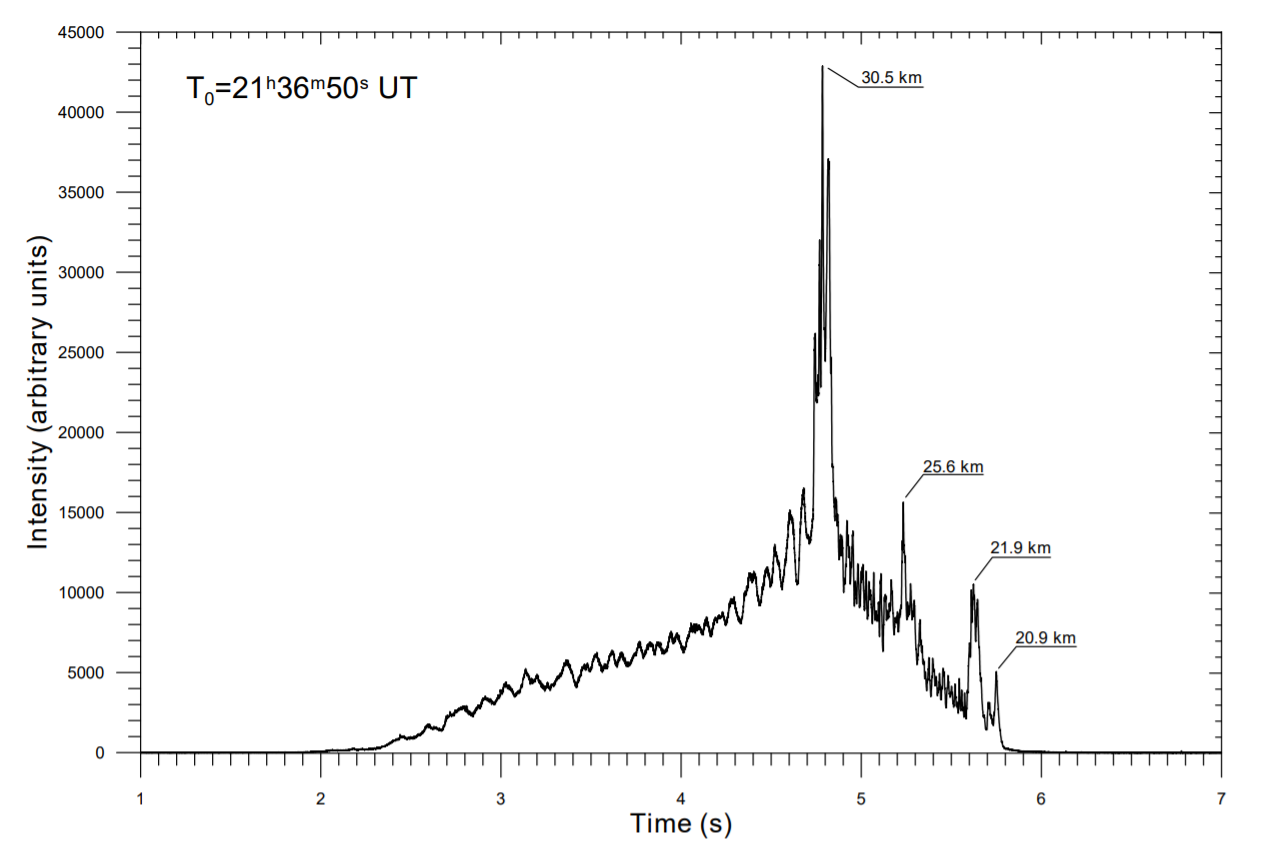}
\caption{The light curve of the Stubenberg meteor from radiometer records. It shows the estimated heights for each fragmentation based on local maxima in the lightcurve and the known trajectory of the fireball from digital camera records \citep{Spurny2016, Borovicka2020}.}
\label{lightcurve}
\end{center}
\end{figure*}

Stubenberg was a meteorite-producing fireball widely observed over Austria, Germany, and the Czech Republic on March 6th, 2016, 21:36:50.495 UTC. The trajectory and lightcurve were precisely recorded by the Czech stations of the European Fireball Network and analyzed from photographic and radiometric data \citep{Spurny2016, Borovicka2020}. The fireball parameters based on these records are shown in the Table \ref{tab:StubenbergTrajectory} and the fireball lightcurve is shown in Figure \ref{lightcurve}. \corrb{A map of the fireball ground track and nearby infrasound and seismic stations is shown in Figure \ref{groundtrack}}. \corrb{ The camera recordings of light production per unit trail length provide high fidelity records of a meteor's energy deposition \citep{Ceplecha1998}. Such lightcurve energy estimates have been validated in several previous studies of meteorite producing fireballs, for which material has been recovered and hence are from events with well constrained initial masses \citep[e.g.][]{Borovicka2013a, Spurny2020}.}

From maxima in the lightcurve, Stubenberg shows clear fragmentation points at the following heights: 20.9, 21.9, 25.6 and 30.5 km with the major fragmentation point at 30.5 km. Stubenberg was chosen as a case study for \texttt{BAM} as the fireball trajectory occurs less than 100 km ground range from a large infrasound array (IS26 - Freyung) and the area is densely covered by seismic stations (more than a dozen within 150 km of the fireball).\par
\begin{figure*}[ht!]
\begin{center}
\includegraphics[width=\linewidth]{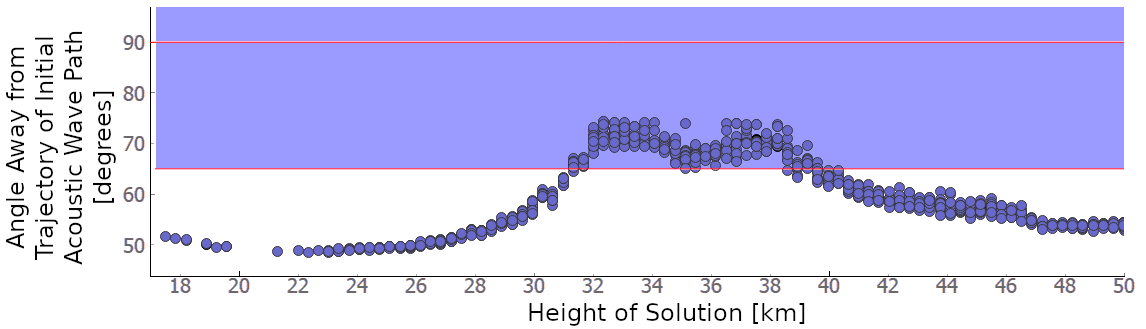}
\caption{The angle of the initial launch ray relative to the fireball trajectory vector to station GR-I26H1 as a function of height for station I26. The height which  is closest to specular (which is at $90^{\circ}$ - red horizontal line) is between roughly 32 and 39 km. We consider probable ballistic arrivals as angles within $\sim 25^{\circ}$ of $90^{\circ}$ \corrb{(the tolerance shown by the blue highlight)}, which indicates it may be a ballistic arrival.}
\label{staffangle}
\end{center}
\end{figure*}
\begin{figure*}[]
 \vspace*{2mm}
    \begin{center}
    \includegraphics[width=\linewidth]{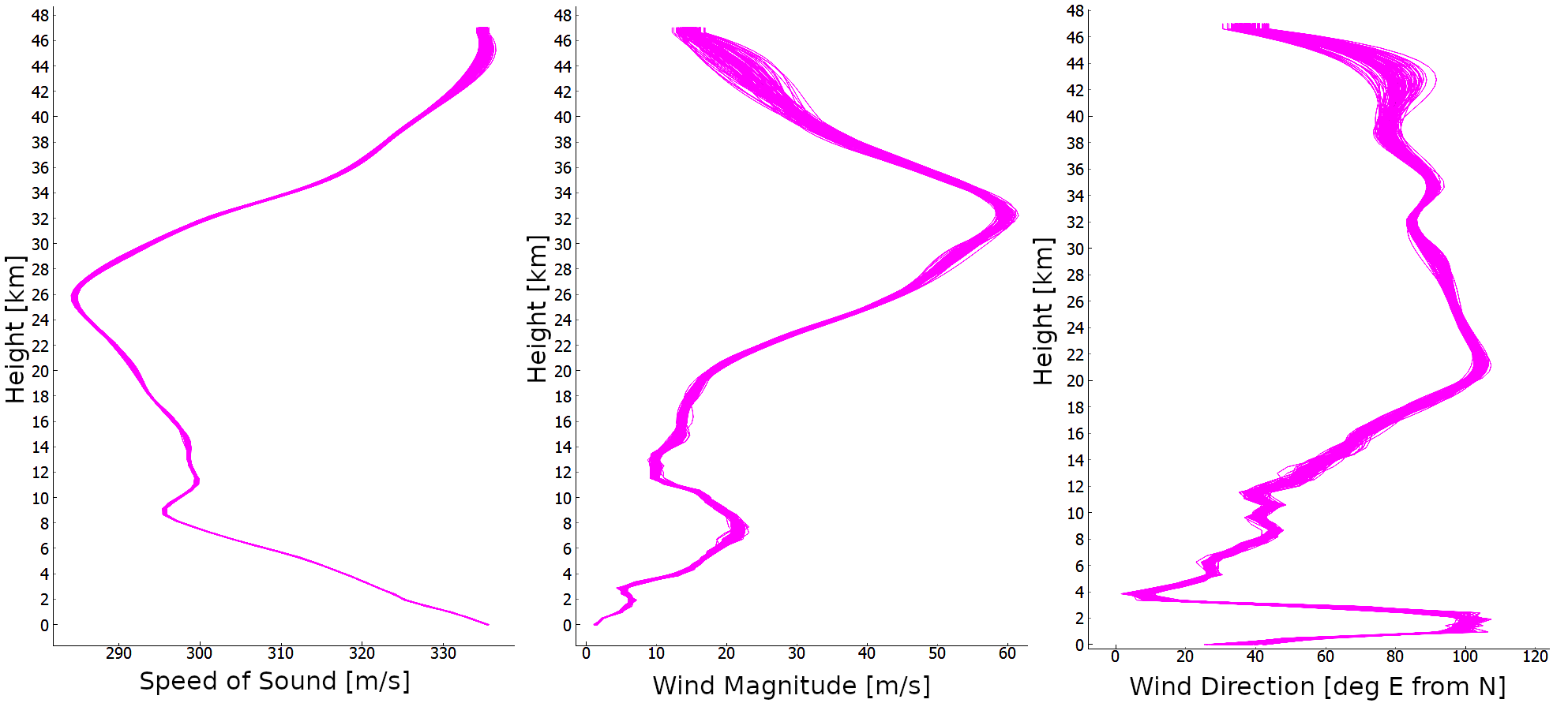}
  \caption{Model sound speed, wind speed, and wind direction profile above the geometric landing point of the Stubenberg event. The pink lines represent perturbed realizations. Note that we use the convention of ERA5 by defining the wind direction as the direction the wind is blowing towards.}
  \label{atmperturb}
  \end{center}
\end{figure*}

From the measured portion of luminous flight, the straight-line trajectory was extended to the geometric landing point located at 13.07393°N, 48.30790°E. The ground track of this fireball \corrb{and the nearby seismic and infrasound stations } are shown in \corrb{Figure \ref{groundtrack}}. The acoustic ground footprint of the fireball when winds and the nominal atmospheric temperature profile are shown in Figure \ref{ball_no_wind}. When the Monte Carlo atmospheric perturbation are also included (see Figure \ref{atmperturb}) it was found that the rays within our adopted tolerance of $\pm$ $25^\circ$ could just reach the IS26 array (see Figure \ref{staffangle}) so we expect a ballistic (as well as possible fragmentation arrivals) at IS26. \corrb{Here IS26 is referred to as the entire I26 array, including stations GR-I26H1 to GR-I26H8}. \par

\begin{table}[]
\begin{tabular}{c|c|c}
\multirow{3}{*}{\begin{tabular}[c]{@{}c@{}}Initial\\ Position\end{tabular}} & Latitude {[}°N{]}  & $48.05970$ $\pm 0.00027$  \\ \cline{2-3} 
                                                                            & Longitude {[}°E{]} & $13.10849$ $\pm 0.00014$  \\ \cline{2-3} 
                                                                            & Height {[}km{]}    & $85.923$ $\pm 0.015$      \\ \hline
\multirow{3}{*}{\begin{tabular}[c]{@{}c@{}}Final \\ Position\end{tabular}}  & Latitude {[}°N{]}  & $48.27900$ $\pm 0.00009$  \\ \cline{2-3} 
                                                                            & Longitude {[}°E{]} & $13.07779$ $\pm 0.00006$  \\ \cline{2-3} 
                                                                            & Height {[}km{]}    & $17.194$ $\pm 0.005$      \\ \hline
\multicolumn{2}{c|}{\begin{tabular}[c]{@{}c@{}}Initial Velocity \\ {[}km/s{]}\end{tabular}}      & $13.913$ $\pm 0.011$      \\ \hline
\multicolumn{2}{c|}{\begin{tabular}[c]{@{}c@{}}Time {[}s{]} \\ (UTC)\end{tabular}}               & $21:36:50.495$ $\pm 0.01$ \\ \hline
\multicolumn{2}{c|}{\begin{tabular}[c]{@{}c@{}}Azimuth \\ {[}° from N +E{]}\end{tabular}}        & $354.67$ $\pm 0.03$       \\ \hline
\multicolumn{2}{c|}{\begin{tabular}[c]{@{}c@{}}Zenith \\ {[}° from Vertical{]}\end{tabular}}     & $19.69$ $\pm 0.02$       
\end{tabular}
\caption{Precise trajectory measured by \citet{Spurny2016, Borovicka2020} of the Stubenberg fireball. The local apparent radiant azimuth and zenith distance are shown together with the initial velocity. The time here corresponds to the first sighting of the meteor, at the initial height}
\label{tab:StubenbergTrajectory}
\end{table}

\begin{figure}[ht!]
\begin{center}
\includegraphics[width=\linewidth]{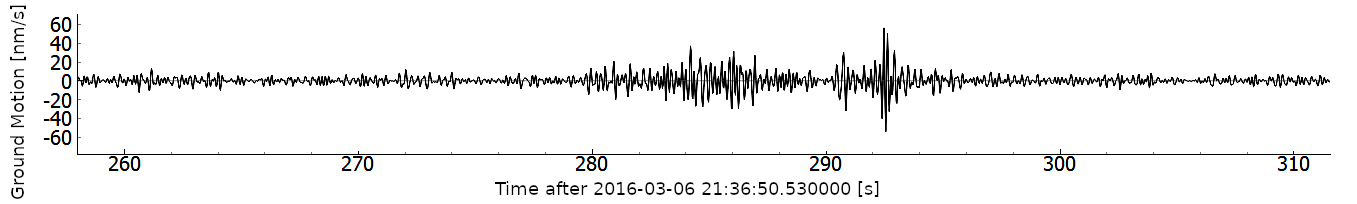}
\caption{ The waveform from seismic station GR-GEC1 in the SHZ channel showing signal from the Stubenberg fireball. The timing is in seconds after the initial optical observation of the fireball. The station is at a ground range of 74 km from the meteor, or 93.36 km in range from 40 km in height. It should be noted that this station detects the same arrival as GR-I26H1, but in a different passband. Station location: 48.84045$^{\circ}$N, 13.70891$^{\circ}$E}
\label{GEC1}
\end{center}
\end{figure}

\begin{figure}[ht!]
\begin{center}
\includegraphics[width=\linewidth]{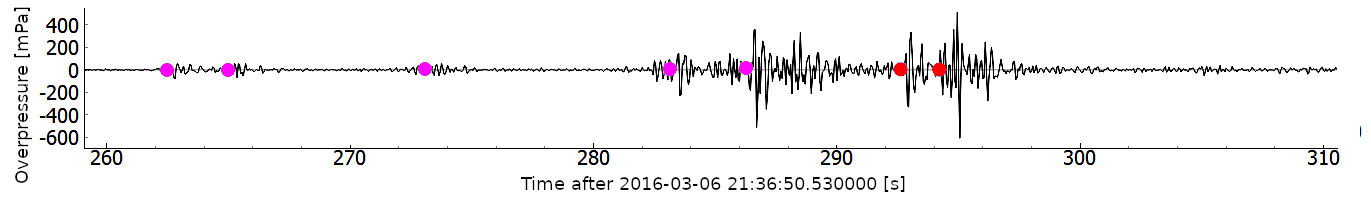}
\caption{Manual picks of discrete arrivals on the GR-I26H1 waveform. The magenta symbols are the probable acoustic fragmentation arrivals, while the red picks are probable ballistic arrivals.}
\label{picked}
\end{center}
\end{figure}

From the acoustic ground footprint shown in Figure \ref{ball_no_wind}, we can identify the stations likely to detect ballistic arrivals. We find that near the IS26 infrasound array the ballistic arrival should be visible. The acoustic signal from the fireball at the \corrb{I26H1 station} (Figure \ref{picked}) is complex and shows several phases. A co-located seismic station (within 900m of the centre of the array) shows a very similar seismo-acoustically coupled signal (Figure \ref{GEC1}). \par \corrb{GR-I26H1 is one infrasound element of the IS26 array, with a broadband pressure sensor in the infrasound regime (channel code BDF). GR-GEC1 is a nearby, short-period, high-gain seismometer (channel code SHZ). These stations are expected to observe the same acoustic signal arrival times, as they are co-located. However, they will likely have different signal amplitudes, since GR-I26H1 will observe the acoustic waves directly, while GR-GEC1 detects air-coupled Rayleigh waves. Further complicating interpretation, GR-GEC1 may also detect precursor ground waves excited by the acoustic waves coupling with the ground at a point distant from the station \citep{Edwards2008}. This demonstrates how seismic and infrasound stations can both augment and complement the data record when performing seismo-acoustic inversions of meteor trajectories.} 
\par


\begin{figure}[ht!]
\begin{center}
\includegraphics[width=\linewidth]{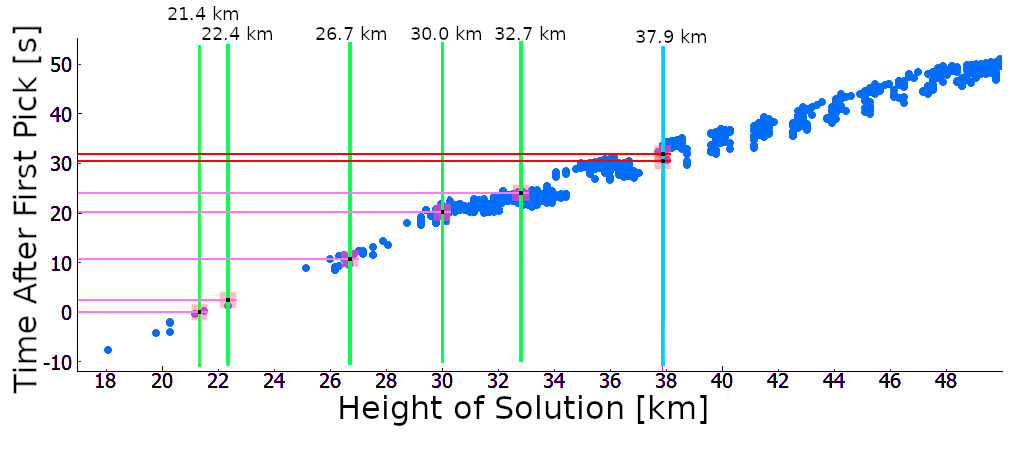}
\caption{The pick timings as used in Figure \ref{picked} (now on the vertical axis) as a function of the corresponding heights of each acoustic source based on the optically-determined fireball trajectory (x-axis). Each blue point represents the timing for a ray-traced acoustic arrival originating at a given height along the fireball trajectory at station GR-I26H1. All Monte Carlo atmospheric perturbations with arrival solutions are shown. The horizontal magenta lines here correspond, in order from bottom to top, the magenta fragmentation-source arrivals in Figure \ref{picked}, from left to right, and the red lines to the red ballistic arrivals. The associated best-estimate for the origin height of each arrival is shown across the top of the plot, with green vertical lines for the fragmentations, and a cyan vertical line for the ballistic arrival.}
\label{staff}
\end{center}
\end{figure}

Figure \ref{picked} shows manual picks on the GR-I26H1 infrasound waveform where potential discrete events are visible which may be generated either by fragmentation events or the ballistic arrival. Here the magenta dots show fragmentation arrivals in chronological order, and the red dots show the probable ballistic arrivals. These identifications are based on the time/height results shown in Figure \ref{staff}. 

In Figure \ref{staff}, the arrival times represented by the magenta dots in Figure \ref{picked} are shown as horizontal magenta lines and translated to height along the optically-determined trajectory based on timing and raytracing (using the mean atmosphere) to GR-I26H1, shown as green vertical lines. Similarly, the specular point (where the ballistic arrival shock should  be generated given the geometry between the station and the fireball velocity vector) along the trajectory as seen from IS26 is shown as vertical cyan line in Figure \ref{staff}. The two horizontal red lines in Figure \ref{staff} represent the timing of the red dots as a function of height along the trajectory in Figure \ref{picked}. \par

It can be seen that the arrival times slowly increase with height until 37.9 km when the specular point is reached. The first four to five picks line up well with the corresponding light curve peaks, which are shown in Figure \ref{lightcurve}. However, the remaining two late picks appear to come from heights higher than the fragmentation points recorded in the light curve. These more closely match the timing/heights expected from the ballistic shock. We interpret these late arrivals are being from the cylindrical shock, though the temporal proximity to the main fragmentation point may indicate some blending between the ballistic and main fragmentation arrival. 

\begin{figure*}[ht!]
\vspace*{2mm}
\begin{center}
\includegraphics[width=\linewidth]{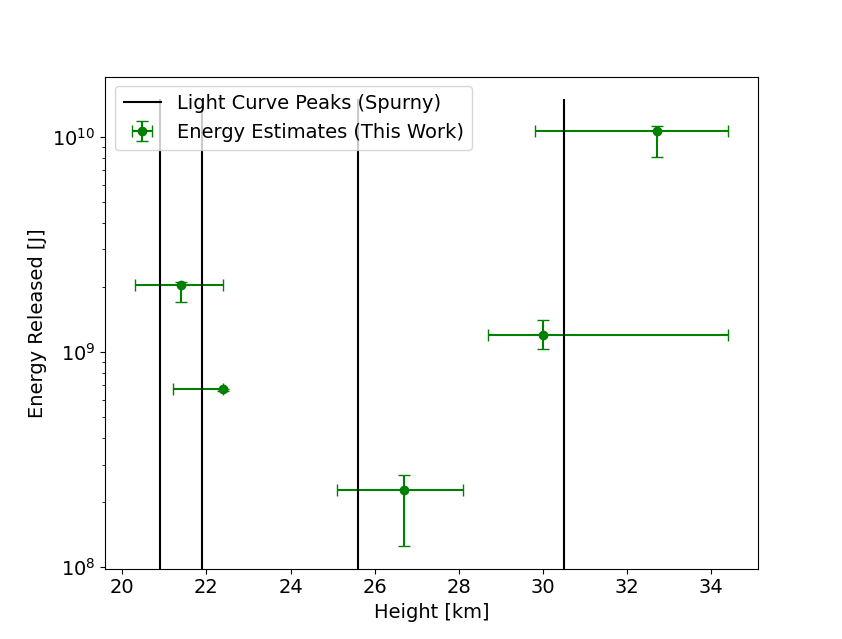}
\caption{The estimated energy per fragmentation event for the Stubenberg fireball determined from measured overpressure. Each fragmentation point is shown superimposed with the heights of fragmentation obtained from the light curve of the fireball (vertical lines). The uncertainty of the height is set by our threshold of requiring a ray arrival at the station $\pm$ 3 seconds (about 1 km) from the observed arrival time}
\label{energyestimates}
\end{center}
\end{figure*}

\begin{figure}[ht!]
\begin{center}
\includegraphics[width=\linewidth]{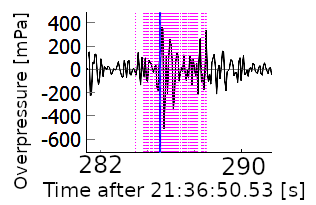}
\caption{The timing of arrivals as expected for I26 from a sample fragmentation point at a height of 32.7 km of the Stubenberg fireball using both the nominal atmosphere and 100 perturbed atmospheric realizations. The dotted lines represent the perturbed atmosphere arrivals while the solid line is the nominal atmosphere. Note that the spread in this arrival is on the order of a several seconds.}
\label{examplespread}
\end{center}
\end{figure}

To better estimate the full range of acoustically accessible heights as seen from IS26 we examine the effects of the Monte Carlo atmosphere perturbations on the acoustic arrival times. Figure \ref{staff} shows that different realizations (multiple dots for each height) produce a range of accessible heights at IS26. The perturbation variables were calculated from the ensemble members provided by the ERA5 data, and are shown below in Figure \ref{atmperturb}. The perturbations were made on the raw u- and v-components of the winds \corrb{(the East-West and North-South  components, respectively)}, and the temperature.  These uncertainties in height are reflected in the uncertainties in source heights given in Table \ref{tab:energyestimates}. \par

 As expected, each atmospheric perturbation was found to produce a slightly different arrival time from that determined using the nominal atmosphere. For the case of the Stubenberg fireball and the geometry to IS26, as shown in Figure \ref{staff} by the spread in points, this typically was of order 2-3 seconds. The range in timing between the latest and earliest arrival represents the uncertainty due to the uncertainty in the atmospheric model. From the timing spread of order 3 seconds in Figure \ref{staff}, the corresponding uncertainty in height is identified  and used in Table \ref{tab:energyestimates}.
 
 As each height corresponds to a different ambient pressure, atmospheric mass density, and radial range from the station, the resulting uncertainty in the energy calculation is dominated by the uncertainty in the height. \par
Since ray paths using perturbations take a slightly different trajectory to the station than using the nominal atmosphere, it is possible to see a perturbation arrival from a height at which a nominal arrival was not observed. \corrb{This is observed because a perturbed atmosphere will have different shadow zones than the nominal atmosphere}. For example, compare Figures \ref{examplestaff} and \ref{staff}, which show the same arrival for GR-I26H1. Figure \ref{examplestaff} shows only 10\% of perturbed arrivals in Figure \ref{staff}, however, the perturbed atmospheres show arrivals from heights below 33.85 km, which are not captured in Figure \ref{examplestaff}. \par
Unfortunately, due to a lack of stations detecting a clear ballistic arrival, the trajectory of Stubenberg could not be inverted from acoustic data alone. However, we have tested the trajectory inversion routine in \texttt{BAM} using other well documented fireballs, such as the Carancas fireball in 2007, and found results similar to published values in \citet{Brown2008, Pichon2008}. 

\subsection{Ballistic Energy Calculation}
The ballistic energy deposited per unit path length was found using the period of the ballistic arrival at IS26 and the forward modelling procedure described in Section \ref{sec:balen}. The result is shown in Table \ref{tab:ballEnergies} and compared to the equivalent blast radius computed form optical records (final column). The ballistic arrival at GR-I26H1 had a dominant period of 0.74 $\pm$ 0.02 s. From ray tracing, the source location on the fireball trajectory of the ballistic wave was found to be approximately $48.2105^{\circ}$N, $13.0873^{\circ}$E, $37.95$ km. The photometrically estimated equivalent meteoroid diameter was about 0.65 m based on the photometric mass of 450 kg \citep{Borovicka2020} and a density of 3129 kg/m$^3$, consistent with the recovered meteorites. The Mach number of the fireball was measured by optical instruments to be $38.68 \pm 0.04$. The acoustic energy, assuming weak shock propagation dominates to the ground as found to be most accurate for smaller fireballs by \citet{SilberBrown2015}, agrees within uncertainty with the photometrically estimated total energy.

\begin{table}[]
\centering
\begin{tabular}{c|c|c|c}
Method & \begin{tabular}[c]{@{}c@{}}Assumed \\ Linear \\ Period\\ T = 0.74 \\ $\pm$ 0.02s\end{tabular} & \begin{tabular}[c]{@{}c@{}}Assumed \\ Weak-Shock\\ Period\\ T = 0.74 \\ $\pm$ 0.02s\end{tabular} & \begin{tabular}[c]{@{}c@{}} $R_0  = \left( \frac{E}{P} \right)^{1/2}$ \\ $\approx 0.742 d_m M$ \end{tabular} \\\hline
\begin{tabular}[c]{@{}c@{}} Blast \\ Radius: \\$R_0$ {[}m{]} \end{tabular} & $20.8^{+0.9}_{-0.8}$ & $19.5^{+0.8}_{-0.8}$ & $18.66^{+0.01}_{-0.03}$ \\ \hline
\begin{tabular}[c]{@{}c@{}}Energy per \\ path length:\\ $E/L$ {[}kJ/m{]}\end{tabular}
  & $172^{+13}_{-13}$ & $151^{+12}_{-12}$ & $138.2^{+0.2}_{-0.5}$ \\ \hline
\begin{tabular}[c]{@{}c@{}}Kinetic \\ Energy: \\ $K$ {[}GJ/m{]}\end{tabular} & $60^{+9}_{-6}$ & $50^{+6}_{-6}$ & $43.6^{+0.1}_{-0.2}$
\end{tabular}
\caption{The blast radii and the corresponding energy for the Stubenberg fireball computed from the dominant period of the ballistic wave arrival at IS26. The first column assumes that the acoustic wave travelled as a linear wave after the transition altitude is reached (see \cite{Silber2019} for a discussion), while the second column assumes purely weak-shock propagation to the ground, believed to be a more accurate estimator for total energy for small fireballs as shown by \citet{SilberBrown2015}. The final column shows the expected blast radius using Equation \ref{blastradiiD} and the known Mach number together with the estimated diameter from optical records for comparison. Values used are $d_m = 0.65$ m, $M = 38.68 \pm 0.04$, $m = 450$ kg, $P = 396.858$ Pa at 37.95 km. Uncertainties in the blast radius due to atmospheric perturbations were found to be negligible compared to the measurement uncertainty in the periods.}
\label{tab:ballEnergies}
\end{table}

\begin{figure}[]
  \centering
  \begin{minipage}[b]{0.95\linewidth}
    \includegraphics[width=\linewidth]{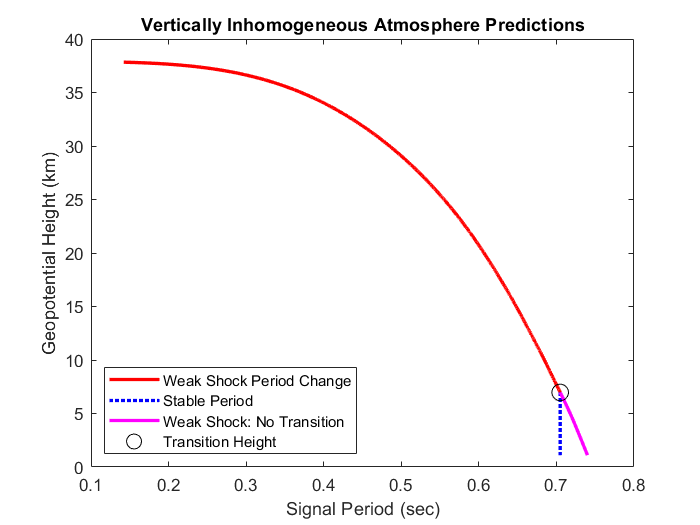}
  \end{minipage}

  \caption{The change in the period of the ballistic wave, as it propagates from the meteor source height of 37.9 km to the infrasound station GR-I26H1. Both the weak-shock and an assumed transition from weak to linear regimes are shown following the approach of \citet{Revelle1974}.}
  \label{matlab}
\end{figure}

Figure \ref{matlab} shows the predicted period at the ground for solutions showing both a linear wave transition and a purely weak-shock path as a function of height. The blast radii found in Table \ref{tab:ballEnergies} were set as initial conditions as forward modelling shows these produced the best fits to the known period and maximum amplitude at the station height, following the methodology of \citet{SilberBrown2015}. It was found that the blast radius derived from the model to match the maximum amplitude were very different to the period approach or the known optical results. This is a similar finding to \citet{SilberBrown2015} who found that for ballistic arrivals, the blast radii estimated from periods were more robust than amplitude-derived values; as a result, we use only period-based blast radii and corresponding energies for this metric in \texttt{BAM}. 

\subsection{Fragmentation Energy Calculation}
\begin{figure}[ht!]
\vspace*{2mm}
\begin{center}
\includegraphics[width=\linewidth]{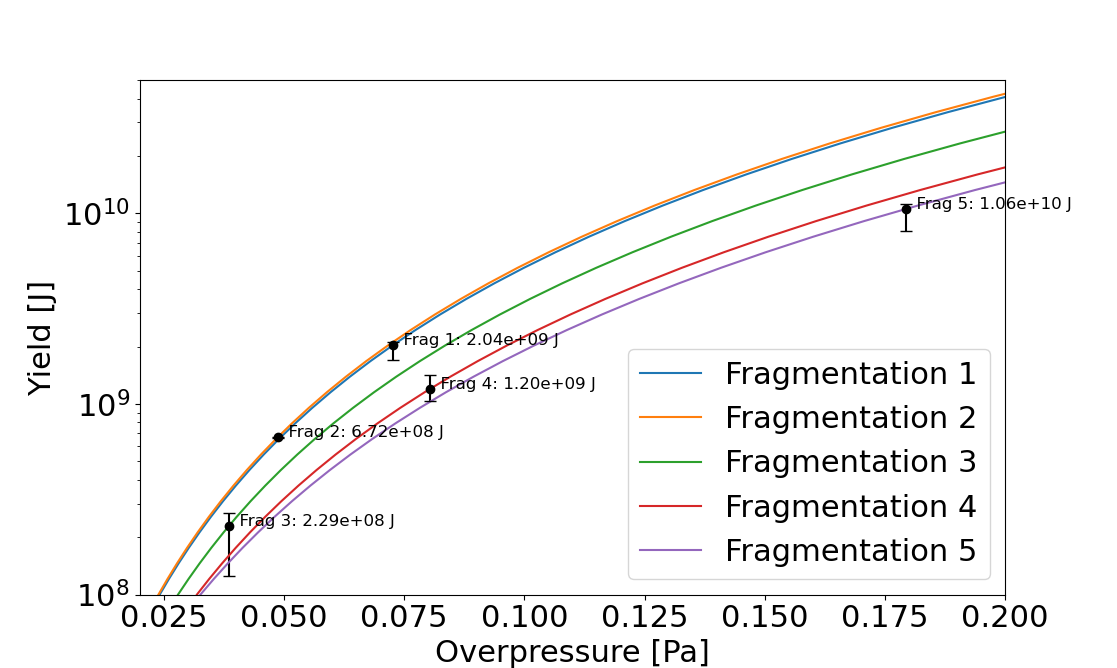}
\caption{Yield vs. overpressure curves of the Stubenberg fragmentation data from I26 measurements. The points on each line represent the overpressure measurement at the station GR-I26H1, and are ordered by ascending heights along the trajectory: Frag 1 - 21.4 km, Frag 2 - 22.4 km, Frag 3 - 26.7 km, Frag 4 - 30.0 km, Frag 5 - 32.7 km.}
\label{yieldcurve}
\end{center}
\end{figure}

The estimated fragmentation heights based on timing arrivals at I26 were extracted from Figure \ref{staff}. These are shown in Table \ref{tab:energyestimates}. The attenuation-corrected relation between spherical source yield and measured overpressure at IS26 for each fragmentation is shown in Figure \ref{yieldcurve}. The heights in the table include the best fitting acoustical fragmentation height and uncertainty as taken from Figure \ref{staff}. For example, the first arrival has a best fit timing from a fragmentation at 21.4 km; however, source heights between 20.3 km and 22.4 km are also possible, within our ray-trace tolerance and including atmospheric perturbations. Figure \ref{energyestimates} shows the corresponding energies of each fragmentation, with vertical lines showing the heights of fragmentation based on maxima in the light curve.

\begin{table}[]
\centering
\scalebox{0.9}{
\begin{tabular}{c|c|c|c|c|c}
\begin{tabular}[c]{@{}c@{}}Height \\ {[}km{]}\end{tabular} & \begin{tabular}[c]{@{}c@{}}Over-\\ pressure \\ {[}Pa{]}\end{tabular} & \begin{tabular}[c]{@{}c@{}}Ambient \\ Pressure \\ {[}kPa{]}\end{tabular} & \begin{tabular}[c]{@{}c@{}}Air \\ Density \\ {[}g/m$^3${]}\end{tabular} & \begin{tabular}[c]{@{}c@{}}Range \\ {[}km{]}\end{tabular} & \begin{tabular}[c]{@{}c@{}}Energy  \\ {[}GJ{]}\end{tabular} \\\hline
$21.4^{+1.0}_{-1.1}$ & 0.073 & $4.5^{-0.7}_{+0.8}$ & $70 \pm 10$ & $82.8^{+0.5}_{-0.5}$  & $2.04^{+0.08}_{-0.34}$ \\
$22.4^{+0.0}_{-1.2}$ & 0.049 & $3.9^{-0}_{+0.8}$ & $60^{-10}_{+0}$ & $83.3^{+0.8}_{-0.7}$  & $0.67^{+0.00}_{-0.01}$  \\
$26.7^{+1.4}_{-1.6}$ & 0.039 & $2.0^{-0.4}_{+0.6}$ & $30^{-6}_{+9}$ & $85.2^{+0.4}_{-0.8}$  & $0.29^{+0.10}_{-0.04}$  \\
$30.0^{+4.4}_{-1.3}$ & 0.080 & $1.2^{-0.6}_{+0.3}$ & $18^{-9}_{+4}$ & $87.8^{+1.3}_{-0.7}$  & $1.2 \pm 0.2$ \\
$32.7^{+1.7}_{-2.9}$ & 0.179 & $0.8^{-0.2}_{+0.4}$ & $12^{-3}_{+7}$ & $89.2^{+1.2}_{-0.6}$ & $10.6^{+2.5}_{-0.6}$  \\

\end{tabular}}
\caption{The heights of major fragmentation points derived from Figure \ref{picked} represented by the magenta pick points. The overpressure was taken from the waveform observed at GR-I26H1, after filtering the data between 0.1 - 2 Hz. The ambient pressures and air mass densities at the respective heights were obtained through the L137 model from ECMWF \citep{Dee2011}. The energies were computed from the corresponding equations in the text, specifically Equation \ref{g_inv}. The uncertainties represent the range in possible ray tracing heights with a tolerance of 3 seconds (1 km) of the station. The nominal value for the height represents the optimal arrival height. The total combined energy from all fragmentations is $1.47^{+0.28}_{-0.12} \times 10^{10}$ J}
\label{tab:energyestimates}
\end{table}

\section{Discussion} \label{discussion}


\subsection{Stubenberg Energy Analysis}
For our Stubenberg case study, we found that the atmospheric uncertainties only slightly change the arrival times to IS26, typically by no more than a few seconds (see Figure \ref{examplespread}). This emphasizes that stations close to a fireball path always produce the most accurate records. For some stations in our case study, perturbed arrivals were predicted where no arrivals were recorded and vice versa. This underscores both the effects uncertainty in the atmosphere can have in interpretation of fireball acoustics and the role scattering/diffraction may play in propagation.  Increasing the number of available perturbations, using on the order of hundreds rather than ten, would better define the uncertainty of arrivals at each station. \par

The height uncertainties shown in Figure \ref{energyestimates} indicate that for Stubenberg and IS26, some of the acoustic fragmentation signals were blended together in the acoustic waveform and should be grouped together. For example, if fragmentations 4 and 5, at 30.0 km and 32.7 km respectively, were grouped together, the data would better match the light curve. Since these fragmentations are from the same general feature shown in the acoustic waveform (see Figure \ref{picked}) and the corresponding light curve feature is broad, it is possible that these two fragmentations are actually one extended event.\par
For Stubenberg, the best estimate for the initial photometric mass was 450 kg. Using the measured speed of 13.9 km/s, the total kinetic energy of the meteoroid is found to be about $4.36 \times 10^{10}$ J. This is in good agreement with the energy computed using the period-only estimate of the ballistic-shock  of $5 \times 10^{10}$ J, particularly considering model uncertainties such as luminous efficiency. In this sense, we may conclude that the Stubenberg meteoroid was still largely a single body at the IS26 specular height of 38-40 km. This is reasonable as this height is above the point of first major fragmentation. 

The total fragmentation energy was found to be $1.47^{+0.28}_{-0.12} \times 10^{10}$ J. This is about a third of the total photometric energy and is therefore physically reasonable. The relative magnitude of the fragmentations, particularly the final two fragmentations combined, also qualitatively match the relative brightness of the fragmentations from the light curve. This represents the first estimates of both total fireball energy and energy partitioned into fragmentation events from purely acoustic measurements. For the Stubenberg case study, the agreement in energy with the optical data and internal consistency in energy estimates between total energy from ballistic signals and fragmentation energy provides a basic validation of our analytic method of bolide fragmentation energy estimation. \par

To simplify the model, it has been assumed in all sections that the fireball velocity was constant. Actual fireballs decelerate rapidly at heights where wave release points would be found. We justify this approximation in this work because the acoustic travel time (on the order of $10^2$ s) and its atmospheric uncertainty (on the order of 1 - 10 s) are much greater in time than the difference in timing of a wave release point for a constant and decelerating fireball. In the future, we hope to include fireball deceleration into \texttt{BAM}.

\subsection{Stubenberg Velocity Analysis}
Finally, using the Stubenberg trajectory and station geometry, we can examine the question of the magnitude of the timing offsets to be expected based solely on the finite velocity of the fireball. In particular,   the sensitivity the entry angle plays in the expected precision of acoustic fireball speeds.

To perform this analysis, the station locations from the Stubenberg fireball analysis were used together with synthetic fireball trajectories for zenith angles of $5, 25, 45, 65,$ and $85$ degrees while keeping the radiant azimuth fixed at the observed value of 354.67$^{\circ}$. Recall that the actual Stubenberg fireball has a zenith angle of ~20 degrees. The wave release point, and the travel time of the acoustic ray to each station for each trajectory were found using the ray-tracing software in \texttt{BAM} with 100 atmospheric perturbations. The offset with atmospheric uncertainty was calculated for each station with each zenith angle. \par
In total 18 stations from the Stubenberg event were used in this analysis. The criteria for a station was that:
\begin{itemize}
  \item It was sufficiently close to the event ($<$ 200 kms)
  \item It was not part of a cluster (only one station from each cluster was used)
  \item It had arrivals for at least 2 different model trajectories
\end{itemize}

\begin{figure}[ht!]
\vspace*{2mm}
\begin{center}
\includegraphics[width=\linewidth]{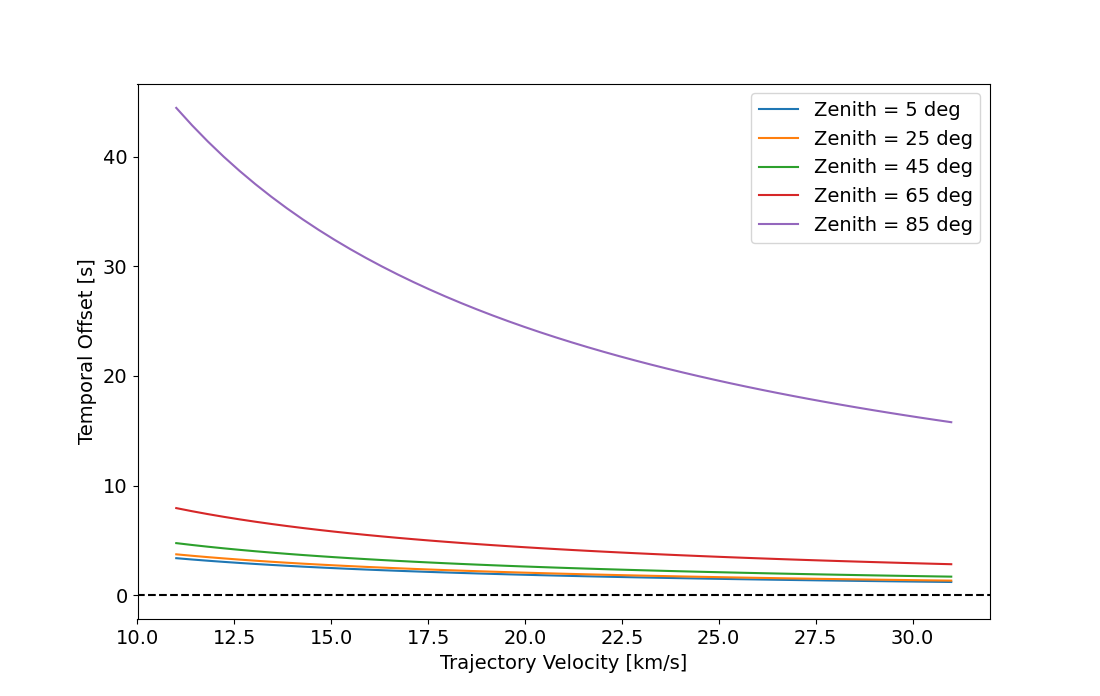}
\caption{An example showing the relative arrival times of the ballistic acoustic wave at seismic station BW-FFB1, as calculated from Equation \ref{theo_eq} using the nominal atmospheric model for Stubenberg. As expected, the model trajectory with zenith angle of $85^{\circ}$ shows much larger timing offsets due to the finite fireball speed than the other trajectories. The trajectory position had to be changed for a zenith angle of $85^{\circ}$, since the station moved outside of the boom corridor, but since only relative times are important for resolving (the steepness of the slope), this movement will have no affect on the result other than increasing all arrival times by a constant amount.}
\label{FFB1exampletheo}
\end{center}
\end{figure}

Figure \ref{FFB1exampletheo} shows the offset-velocity plot for seismic station BW-FFB1. The general trend for all stations is as expected from Equation \ref{theo_eq}. Temporal offsets are greater at lower velocities, which is shown by the $\Delta t \propto \frac{1}{v}$ relation. Higher zenith angles, in most cases, had a larger difference in temporal offset from the higher velocities to the lower velocities. This demonstrates that even modestly shallow zenith angle of order 65 degrees and slow fireballs may show timing variations due to finite speed of more than several seconds which may allow for velocity estimates. \corrb{Very shallow entry fireballs can in principle have relatively high precision measurements made of fireball velocity based purely on timing differences for proximal stations. In practice, the combination of a deeply penetrating fireball with shallow entry near a dense seismic network while possible is rare .} \par 

\begin{figure}[ht!]
\vspace*{2mm}
\begin{center}
\includegraphics[width=\linewidth]{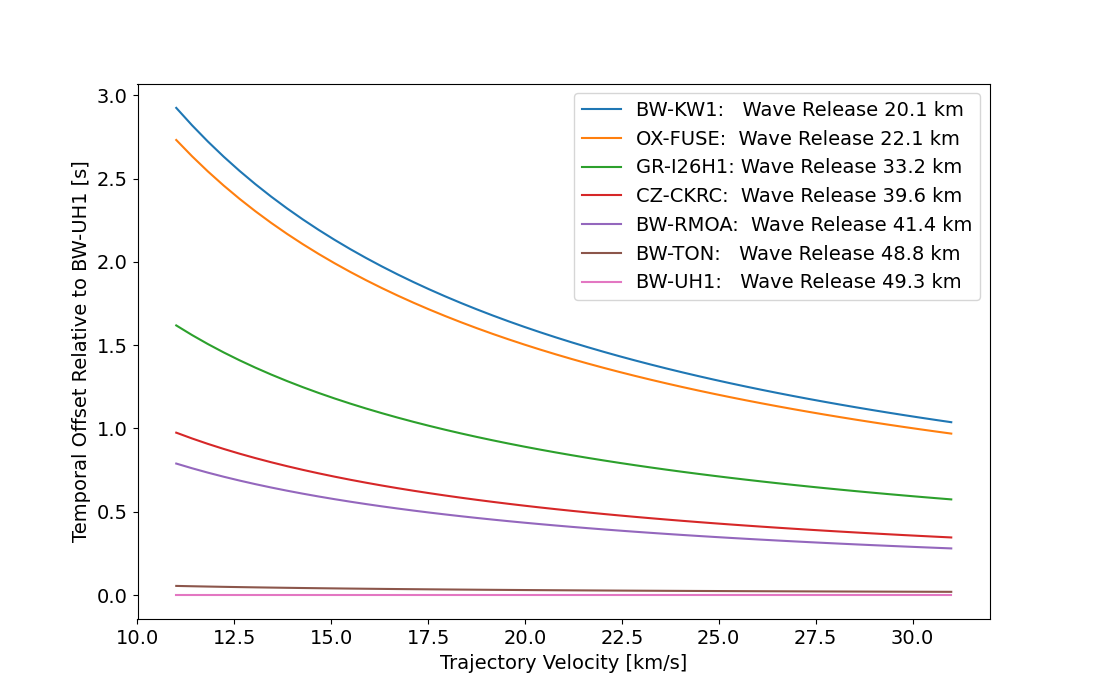}
\caption{The relative arrival times of the ballistic acoustic signal from a fireball with a zenith angle of 25$^{\circ}$ at various stations on the ground, normalized to the timing at BW-UH1. Note that the further away the wave release point is from 49.3 km, the larger the offset in time.}
\label{relativeheights}
\end{center}
\end{figure}

Having a greater separation in wave release points (i.e. different specular points), and effectively a larger $\Delta h$, results in a more resolvable velocity. Figure \ref{relativeheights} shows a subset of stations with their offset measured relative to that of BW-UH1. This shows that using stations having widely separated wave release heights will always produce better fireball velocity estimates. \par
\begin{figure}[ht!]
\vspace*{2mm}
\begin{center}
\includegraphics[width=\linewidth]{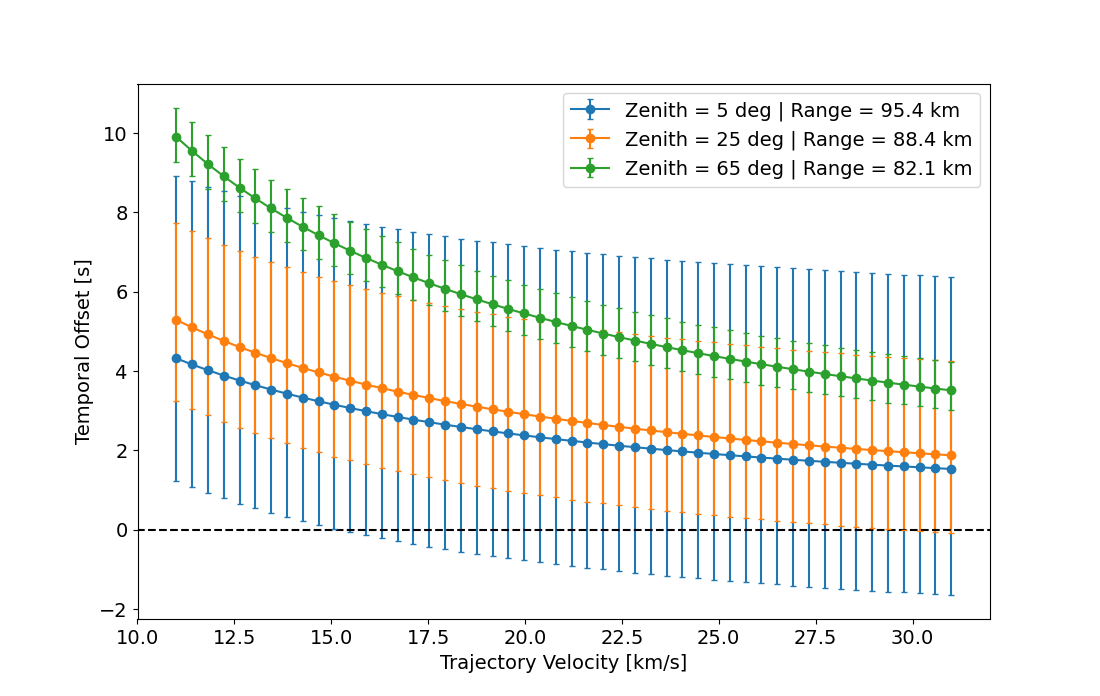}
\caption{The relative arrival times of a ballistic wave at station GR-I26H1 as a function of velocity for different entry angles using the Stubenberg fireball azimuth and station distribution. The uncertainty bars here show the absolute range in arrival times from a spread of 100 atmospheric realizations. Note that the zenith angle, and therefore the range most affects the uncertainty in arrival times.}
\label{I26H1_range_perts}
\end{center}
\end{figure}

It was found that the timing spread using different atmospheric realizations, in general, increased with the range from the wave release point as expected. Figure \ref{I26H1_range_perts} shows the spread due to differing travel times associated with the realizations. The further the acoustic wave must travel through the atmosphere, the more the timing spread in the final arrivals. Typically, this means that grazing fireballs tend to be more likely to produce measurable speeds since they are more likely to have a wave release point closer to a given station. In general, the velocity is not measurable at ranges in excess of 100 km due to the variability of the atmosphere. This again emphasizes the importance and value of near field acoustic measurements of fireballs.  \par

\section{Conclusions}
We have described details of a new Bolide Acoustic Modelling (\texttt{BAM}) computer program, designed to measure fireball energetics, fragmentation points and trajectories. A primary goal of \texttt{BAM} is to provide a framework to separate fireball ballistic acoustic arrivals from those produced by fragmentation events. In particular, we have presented a new method for estimating fragmentation energy from fireballs using the acoustic overpressure measured at the ground as part of \texttt{BAM}. The method assumes knowledge of the location and height of the fragmentation point. 

We have validated this approach in a case study of the Stubenberg meteorite producing fireball. We focused on Stubenberg for calibration as it was well documented by calibrated fireball cameras which produced optical records to independently estimate the total photometric mass/energy of the fireball and which was further constrained by the recovery of a meteorites.  Stubenberg had a well defined late in flight fragmentation episode and was proximal to an infrasound station, making it ideal for our validation requirements. While Stubenberg may be the best available calibration event, we intend to seek other large fireballs with independent trajectory and energy deposition measurements which occurred near infrasound stations for further validation.

Combining the scaled distance model through an atmosphere as outlined by \citet{KinneyGraham1985} and appealing to Sachs Scaling with the attenuation factors shown by \citet{Reed1972, Reed1972a} and further developed by \citet{Latunde2013}, we were able to estimate realistic acoustic attenuations between the IS26 infrasound station and the Stubenberg fireball to estimate energy release per fragmentation episode. \par
Assuming a weak-shock propagation to the surface using the methods of \citet{SilberBrown2015}, the ballistic energy for the Stubenberg fireball was found to be approximately $5 \times 10^{10}$ J which compares favorably to the optical estimate of $4.36 \times 10^{10}$ J. \par
The combined fragmentation energy of the Stubenberg event from acoustic data was found to be $1.47^{+0.28}_{-0.12} \times 10^{10}$J, roughly one third of the ballistic or optical total energy. The relative magnitude of each independent fragmentation, specifically the first and last fragmentations, strongly correlated with the relative fragmentation intensities observed in the light curve, providing further validation of our new acoustic technique for measuring fragmentation energies. \par 
We also explored the role of  atmospheric model uncertainties on acoustic estimates for fireballs, using the ensemble atmospheric perturbations now provided by the ECMWF model. Using atmospheric realizations to provide an estimated uncertainty in arrival times of fireball acoustic signals at stations, we found for the Stubenberg case study variances on the order of a few seconds at the closest stations due solely to the atmosphere uncertainty. The difference in arrival times was directly translated to uncertainty in the height of fragmentations and the ballistic wave release points for the Stubenberg fireball, which were found to be on the order of a few kilometers. \par

We also show that measuring fireball velocities using acoustic data alone is not practical in most cases. Only fireballs which are relatively slow and of grazing incidence occurring proximal to dense seismic/infrasound networks could have reasonably precisely measured velocities.

\section{Acknowledgements}
This work was supported in part by the NASA Meteoroid Environment Office under cooperative agreement 80NSSC18M0046. PGB also acknowledges funding support from the Natural Sciences and Engineering Research council of Canada (RGPIN-
2016-04433) and the Canada Research Chairs program (grant 950-231930). PS work was supported by the Praemium Academiae of the Czech Academy of Sciences.
The facilities of IRIS Data Services, and specifically the IRIS Data Management Center, were used for access to waveform, metadata or products required in this study. The IRIS DS is funded through the National Science Foundation and specifically the GEO Directorate through the Instrumentation and Facilities Program of the National Science Foundation under Cooperative Agreement EAR-1063471. Some activities of are supported by the National Science Foundation EarthScope Program under Cooperative Agreements EAR-0733069, EAR-1261681.

\clearpage
\newpage
\section*{References}
\bibliography{references.bib}

\appendix

\section{Bolide Acoustic Modelling \texttt{BAM} Code Details} \label{BAMdocs}

\subsection{Overview}
\corrb{Here we present more specific details of our bolide acoustic analysis methodology, including how they are implemented in the \texttt{BAM} software package. The \texttt{BAM} Python code is an open-access code with an online Github repository\footnote{https://github.com/wmpg/Supracenter}, which includes detailed user documentation. } \par
\subsection{Acquiring Station Data in \texttt{BAM}}
\begin{figure}[ht!]
\vspace*{2mm}
\begin{center}
\includegraphics[width=\linewidth]{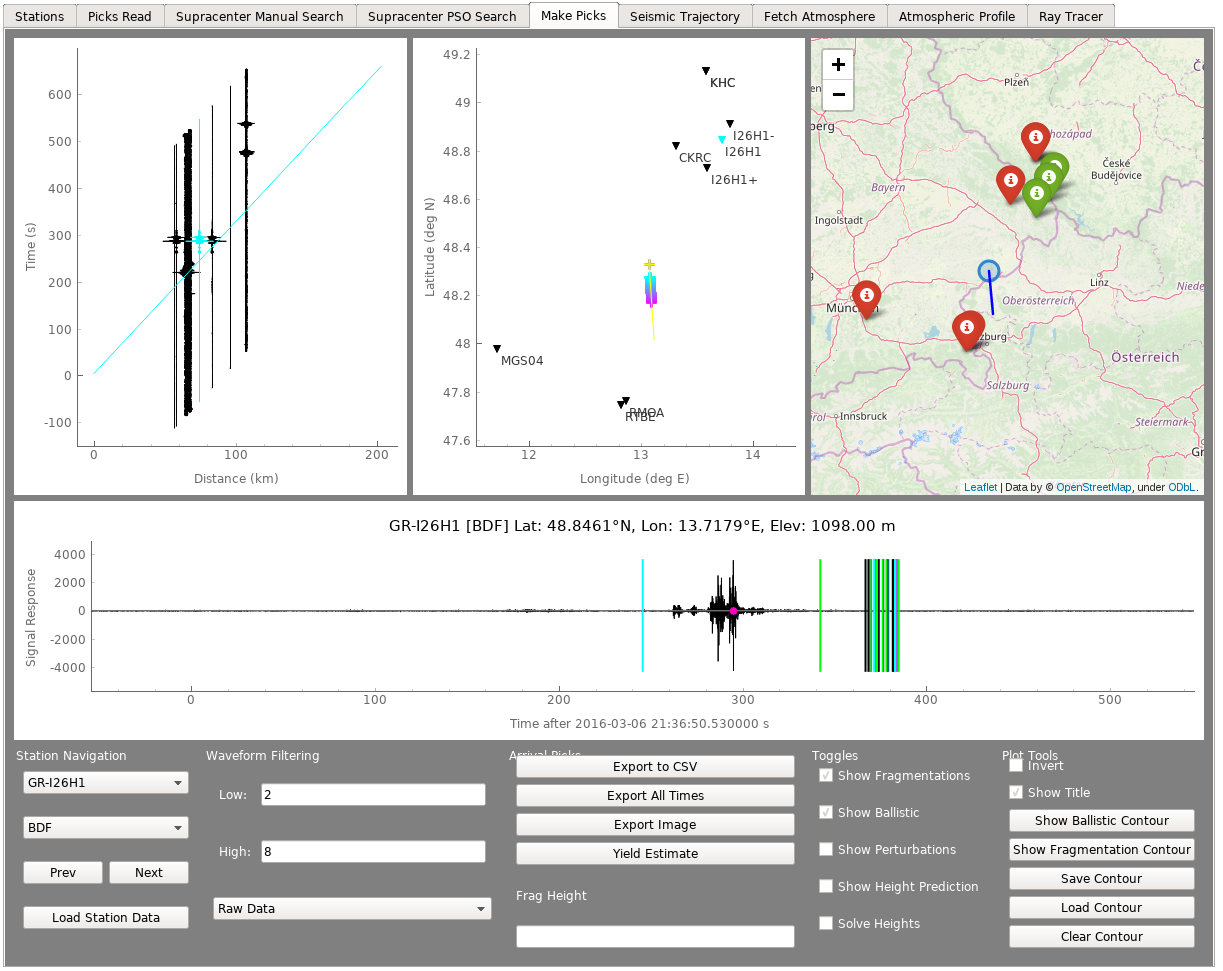}
\caption{A screenshot of the user-interface used to create seismic picks. The user can view, filter, and zoom into the waveform, and create a pick by clicking exactly where specified. Shown here, the user has made a pick by the magenta dot on the waveform (lower plot). The upper left window shows all waveforms sorted by distance away from a reference point. The upper middle and right windows show the user-estimated meteor trajectory as blue lines, surrounded by seismic and infrasound stations denoted as triangles or map points. The bottom user-interface gives user controls for manipulating the waveform. More on the \texttt{BAM} code can be found in the user manual at https://github.com/wmpg/Supracenter.}
    \label{GUI}
\end{center}
\end{figure}

Having the approximate fireball location identified, all available stations (mostly seismic, but some infrasound if available) within 150 km ground range are automatically downloaded from their respective data center using the web services of the International Federation of Digital Seismograph Networks \newline (FDSN)\footnote{https://www.fdsn.org/}. \corrb{These station waveforms are then presented with a map of the area, where the user may scroll through, delete, and annotate the station data as needed }. Figure \ref{GUI} shows a sample image of the user-interface of how waveform picks are made for a chosen station. \corrb{The waveform may be scaled as needed so that the accuracy of the picks are only limited by the sampling rate of the detector.} \par 

The waveforms of each station are viewed and analyzed manually, allowing the user to bandpass filter waveforms, and compare their timings with waveforms from other stations and expected arrival timing envelopes at each station based on the test fireball location. The user manually decides which features of the waveforms are possible fireball associated acoustic arrivals.  \par
Once probable fireball acoustic arrivals are identified at each station and tagged as probable fragmentation or ballistic arrivals, they are imported to either the fragmentation solver, \textit{Supracenter}, or the ballistic solver, \textit{Acoustic Trajectory}. These routines then search for a fragmentation point solution (3D position and time) or overall trajectory orientation (radiant elevation, azimuth, ground location and height) using the L2 norm (sum of squared differences between predicted and observed arrival times) as a cost function to estimate a solution. \par

\subsection{\texttt{BAM} Module : Supracenter}

\corrb{In finding a Supracenter, the arrival times of at least four sufficiently close stations ($<$ 150 km ground range) are required if the time of fragmentation is not known. } A range of points distributed in latitude, longitude, elevation, (first converted into a grid of local coordinates) and source time are used in a search algorithm. The travel time from various x, y, z source ($S$) points are compared to the measured arrival time at each station or detector ($D$), and the point with the minimum total time error is found. \par 
The sub-routine uses the Tau-P ray-tracing algorithm \citep{Garces1998} \corrb{to calculate the arrival times and the ray-paths} . The routine uses wind and temperature values to calculate the total travel times from a range of initial azimuth and zenith angles launched from the fireball. The genetic search algorithm originally used by \citet{Edwards2004} was replaced with a more modern particle swarm optimization search \citep{pyswarm} to improve computation speed and accuracy. \corrb{This Tau-P ray-tracing algorithm has been used previously with vertical weather profiles similar to those taken from ECMWF in the past by \citet{Edwards2003}. It was found that the ECMWF data produces a similar sounding profile to those used in \cite{Edwards2003}}.\par

The algorithm launches rays spaced along an altitude / azimuth grid at the source, and propagates each one until it reaches the ground. The closest ray to the receiver is saved at each step, and a new, denser ray net is formed around that best ray, until the spacing becomes sufficiently small (default of $10^{-10}$ degrees). The eigenray, which is the best estimate of the true raypath from the source to the detector, is found when the three-dimensional range from the end of the ray is within a user-set spatial tolerance of the detector; \corrb{empirically we found that reasonable } default values are 0.3 km ground distance and 3 km vertical height distance. A ray-tracing solution from a given source grid launch point may fail for one of three reasons :
\begin{itemize}
  \item The spacing between test launch angles has become sufficiently small, and no rays propagate within the spatial tolerance of the receiver.
  \item The detector is in a "shadow-zone" and no rays are able to reach the receiver as the temperature/wind structure precludes direct ray paths from the source to receiver.
  \item All rays in the current ray net are refracted upward, and are not sufficiently close to the station.
\end{itemize}
Once eigenrays are found for each station from a given test location the sum of squared time residuals between the model estimate of the ray tracing arrival time and the observed time is used as a cost metric to decide on goodness of fit, as originally employed by \citet{Edwards2004}.\par 
If the acoustic ray propagates to low heights and sufficiently close to \corrb{the detector } but is refracted upward, a signal may still be detected due to diffraction or scattering. Therefore, the three-dimensional spatial tolerance  in \texttt{BAM} is calculated at each step, and if the position of the ray is within a tolerance, (by default set to $330$ m horizontally to represent about 1 second in acoustic travel time, and $2$ km vertically to represent possible acoustic scattering), the ray is still considered an arrival. \par

\subsection{\texttt{BAM} Module : Acoustic Trajectory}

Within \texttt{BAM} the five parameters defining a fireball trajectory (entry angle, entry azimuth and intersection location with the ground $X_0$, $Y_0$, and the time of occurrence which is coupled to the velocity - see Figure \ref{ballistic_model}) providing the best fit are calculated using a particle swarm optimization function \citep{pyswarm}. This finds the residuals in observed-expected arrival times for a given trajectory, and attempts to move this guess so as to minimize this total timing error. \par
\corrb{In finding the optimal trajectory, we use the timings in the expected ballistic arrival. This method uses the same ray-tracing scheme as the Supracenter module, with a few minor differences. First, the trajectory is split into many different supracenters along the trajectory (the \texttt{BAM} default is 100 supracenters between 17 km and 50 km in height on the trajectory). Acoustic waves are propagated from each of these sources to the station. The source with the closest initial launch angle vector being perpendicular to the trajectory vector (while being within $25^{\circ}$ to perpendicular to the trajectory vector) becomes the wave-release point. The timing of arrivals from that wave-release point is then calculated as if it were a supracenter.}

\subsection{Atmospheric Model and Perturbation Scheme}

\texttt{BAM} uses both the temperature and winds from a 37 pressure level, 3-D grid, space with a horizontal resolution of $0.25^{\circ}$. These data are taken from the Copernicus Climate Data Store, specifically from the fifth generation of the ECMWF (European Centre for Medium-Range Weather Forecasts) atmospheric reanalysis (ERA5). This source also provides ensemble member assimilations, which are perturbations added to the mean profiles to indicate atmospheric uncertainty ranges. The spread of these ensemble members are used within \texttt{BAM} to generate Monte Carlo perturbations, providing uncertainty spread in our final solutions due to atmospheric variability. \par

\corrb{With the grid atmospheric data retrieved from ECMWF, a vertical atmospheric sounding profile, similar to those used by \cite{Edwards2003}, is generated by combining vertical sounding profiles from the nearby grid spaces between $S$ and $D$. A straight, 3-dimensional line is taken from $S$ to $D$ to estimate a very rough ray path. The sounding data at each pressure level from the closest latitude/longitude grid point is used in a combined sounding profile, which is then cubic splined to generate a smooth profile } to keep the first derivative of the profile without discontinuities. Such discontinuities cause errors in ray propagation in the ray-tracing algorithm. By default, the cubic spline interpolates 100 points from ground level to the 1 hPa pressure level (approximately 45-50 km). \corrb{Perturbations are generated in a similar method, but using perturbed original sounding data.} 

In the perturbation scheme, the spread of ensemble members \corrb{retrieved from ECMWF } is used as the standard deviation of the error in a nominal atmospheric profile. The realizations are randomly generated by fitting a Gaussian distribution over the nominal profile with the spread defining the standard deviation of the distribution. For each raw variable (temperature, meridional wind, and zonal wind), the value it takes is randomly generated within the Gaussian distribution. The simulations were run using one random seed per realization, one random seed per variable, and a new random seed for each variable and pressure level. It was found that the spread of the arrival times is not significantly dependant on when the seeds are generated (spread changes by less than a factor of 2\corrb{, and the total spread remains on the order of seconds, representing typically $\approx$1\% of total travel time}), however, using different seeds for each level and variable was found to have a larger effect. Here we opt to provide an upper limit to the probable uncertainty by using  different random seeds for each pressure level and variable. \par

\subsection{Energy Estimation in \texttt{BAM}}

Within \texttt{BAM}, the attenuation factor is calculated from Equation \ref{af} in the main text, with the summation over individual pressure layers calculated using the ray-tracing algorithm, \corrb{estimating the attenuation facor at each pressure level } . The frequency, $\nu$, is left as a function of yield, $W$, using Equation \ref{sach}.\par

The geometric correction is found by tracing a ray from the fragmentation source to the station in the nominal and perturbed atmospheres. Four rays are then sent out from the source deviating from the original takeoff angles by an angle $d \theta_{0, nonideal}$, taken here to be $1.5^{\circ}$. The ground area that these four rays produced are then traced back to the source in an isotropic atmosphere, which produces the deviation angle in an ideal system, $d \theta_{0, ideal}$. In this method, $dr$ is kept as a constant, and therefore can be emitted from Equation \ref{rf}:

\begin{equation}
    rf = \sqrt{\frac{d \theta_{0, nonideal}}{d \theta_{0, ideal}}} = \sqrt{\frac{{1.5}^{\circ}}{d \theta_{0, ideal}}}.
\end{equation} 
Each perturbation and nominal arrival refractive factor for a specific fragmentation is then averaged. The perturbation vs. nominal refractive factors are usually found to be similar in value and hence the uncertainty in height dominates the energy uncertainty as compared to any uncertainty in the refractive factor. \par
To find an appropriate $d \theta_{0, nonideal}$, the value must be both small enough to accurately represent the spread of the rays yet not be so small that a single perturbation would cause a ray to refract unphysically. 

\begin{figure}[ht!]
\vspace*{2mm}
\begin{center}
\includegraphics[width=\linewidth]{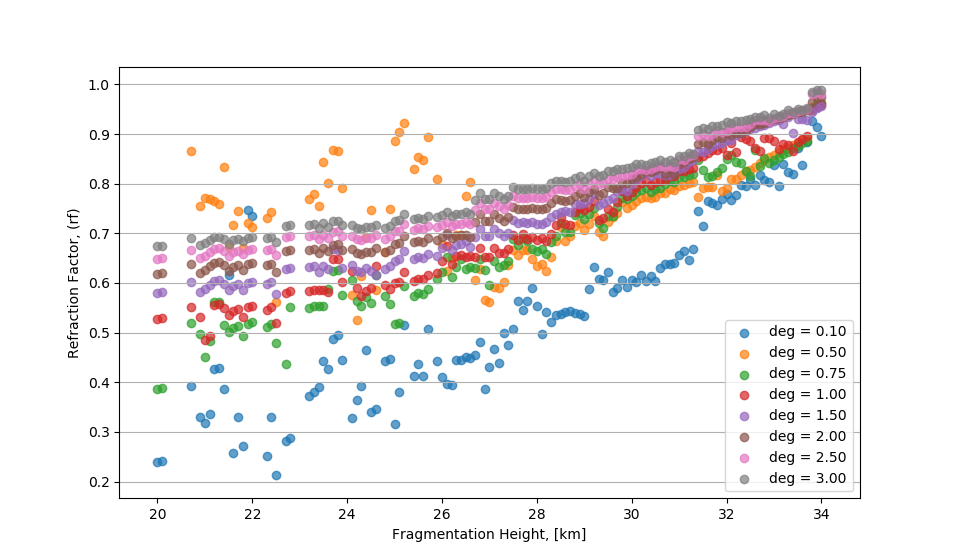}
\caption{The refractive factor to a single station (in this case infrasound station I26) as a function of the height of fragmentation along the trajectory of the Stubenberg meteor. Here different angles are used for $d \theta_{0, nonideal}$. It was found that $d \theta_{0, nonideal} = 1.5^{\circ}$ was the smallest angle tested that showed adequate convergence defined as providing a smooth (as opposed to discontinuous) change as a function of height.}
\label{rfangles}
\end{center}
\end{figure}

Figure \ref{rfangles} shows an example of the corresponding refraction angles as a function of height for our case study of the Stubenberg fireball. Here it is found that $1.5^{\circ}$ is the smallest angle that showed the refractive factor converging to a continuous function of height. \par
The remaining parameters in Equation \ref{g} are atmospheric, and are extracted by \texttt{BAM} from the raw atmospheric data. The constants $k$, $b$, and those with subscripts of $0$ were taken from \cite{Reed1972, Reed1972a}. Values of $f_d$ were calculated from Equation \ref{transmission} and validated by comparison with tables given in \cite{KinneyGraham1985}. The expected $\Delta p_n$ is given as a function of yield, $W$. In our example, this is typically taken from $1 \times 10^8 - 1 \times 10^{12}$ J and Equation \ref{g} is inverted numerically, giving a fragmentation yield estimate. \par

\section{Seismic Station References} \label{stationref}

Table \ref{statreftab} gives the references of the station networks used for our case study of the Stubeberg fireball. For full station lists of each of these networks, see the appropriate Digital Object Identifier (DOI). A network was placed here if at least one station from it was used in the analysis.

\begin{table}[htbp]
\small
\begin{tabular}{|c|l|}
\hline
\begin{tabular}[c]{@{}l@{}} Network \\ Name \end{tabular}& Reference\\ \hline
BW &  \begin{tabular}[c]{@{}l@{}} 	
Department of Earth and Environmental Sciences, Geophysical\\ Observatory, University of Munchen (2001). \textit{BayernNetz}. \\ International Federation of Digital Seismograph Networks. \\ https://doi.org/10.7914/SN/BW \end{tabular} \\ \hline
CZ & \begin{tabular}[c]{@{}l@{}} 	
Institute of Geophysics, Academy of Sciences of the Czech \\ Republic (1973). \textit{Czech Regional Seismic Network}. International\\ Federation of Digital Seismograph Networks. \\ https://doi.org/10.7914/SN/CZ\end{tabular}  \\ \hline
GR & \begin{tabular}[c]{@{}l@{}} 	
Federal Institute for Geosciences and Natural Resources \\ (BGR). (1976). \textit{German Regional Seismic Network (GRSN). }\\Federal Institute for Geosciences and Natural Resources (BGR). \\ https://doi.org/10.25928/mbx6-hr74\end{tabular} \\ \hline
OE &  \begin{tabular}[c]{@{}l@{}} 	
ZAMG-Zentralanstalt Für Meterologie Und Geodynamik. (1987). \\ \textit{Austrian Seismic Network.} International Federation of Digital \\ Seismograph Networks. https://doi.org/10.7914/SN/OE\end{tabular}  \\ \hline
OX & \begin{tabular}[c]{@{}l@{}} 	
OGS (Istituto Nazionale Di Oceanografia E Di Geofisica \\ Sperimentale). (2016). \textit{North-East Italy Seismic Network.} \\ International Federation of Digital Seismograph Networks. \\ https://doi.org/10.7914/SN/OX\end{tabular}  \\  \hline
SL &  \begin{tabular}[c]{@{}l@{}} 	
Slovenian Environment Agency. (2001). \textit{Seismic Network of the} \\ \textit{Republic of Slovenia}. International Federation of Digital \\ Seismograph Networks. https://doi.org/10.7914/SN/SL\end{tabular}  \\ \hline
\end{tabular}
\caption{A list of the seismic networks used in this study. For further information, and for data access, see the network's respective DOI.}
\label{statreftab}
\end{table}

\end{document}